\documentclass{PoS}
\usepackage{amsmath}
\usepackage{amsfonts}
\usepackage{amssymb}
\usepackage{graphicx}
\usepackage{subcaption}
\usepackage{comment}
\title{What does the matter created in high multiplicity proton-nucleus collisions teach us about the 3-D structure of the proton?}

\ShortTitle{What does the 3-D structure of the proton...}
\author{Kevin Dusling\\
        American Physical Society, 1 Research Road, Ridge, NY 11961, USA\\
                Physics Department, Brookhaven National Laboratory, Bldg. 510A, Upton, NY 11973, USA \\
        E-mail: \email{kdusling@mailaps.org}}
        
\author{Mark Mace\\
        Physics Department, Brookhaven National Laboratory, Bldg. 510A, Upton, NY 11973, USA \\
        Physics and Astronomy Department, Stony Brook University, Stony Brook, NY, 11974, USA\\
        E-mail: \email{mark.mace@stonybrook.edu}}
        
\author{\speaker{Raju Venugopalan}\\
        Physics Department, Brookhaven National Laboratory, Bldg. 510A, Upton, NY 11973, USA \\
        E-mail: \email{raju@bnl.gov}}

\newcommand{\Bp}{B_p}
\newcommand{\pmax}{p_\perp^{\rm max}}

\newcommand{\Eq}[1]{Eq.~(\ref{#1})}

\newcommand{\Fig}[1]{Fig.~(\ref{#1})}

\abstract{The study of multiparticle correlations and collectivity in the hot and dense matter created in collisions of heavy-ions as well as those of  smaller systems such as proton--heavy-ion collisions has progressed to the point where detailed knowledge of the three-dimensional structure of the proton is needed to confront experimental data. We discuss results from a simple proof-of-principle initial state parton model which reproduces many features of the data on proton--heavy-ion collisions that are often ascribed to hydrodynamic flow. We outline how this model can be further improved with particular emphasis on missing elements in our understanding of the dynamical spatial and momentum structure of the proton.}

\FullConference{QCD Evolution 2017\\
		22-26 May, 2017\\
		Jefferson Lab Newport News, VA - USA}

\begin{document}

\section{Introduction}
The detection and characterization of the hottest known state of matter created on Earth, the Quark Gluon Plasma (QGP), is a major achievement of experiments on  heavy-ion ($AA$) collisions conducted at RHIC at BNL and the LHC at CERN. The QGP is characterized by its fluid-like nature, which is strongly interacting and possesses the smallest shear viscosity to entropy density ratio found in the nature. A key element in solidifying this consensus about the QGP was the striking observation at RHIC about a decade ago of a novel ``near-side" ridge-like structure in the two particle correlation spectra. This near-side ridge structure was named thus because it was long range in the relative rapidity ($\Delta \eta$) between the particles but collimated  in their relative azimuthal angle ($\Delta \phi\simeq0$)~\cite{Adare:2008ae,Alver:2009id,Abelev:2009af}. The ridge was more pronounced for central (head-on) heavy-ion collisions than in peripheral collisions. In contrast, the peripheral collisions exhibited an ``away-side"  ridge-like structure in the relative rapidity of two particle correlations that are back-to-back in their relative azimuthal angle ($\Delta \phi\simeq \pi$). The away-side ridge structure is greatly diminished with increasing centrality of the heavy-ion collision. 

The systematics of the ridge-like structures in the two particle correlation spectra in $AA$ collisions has a compelling explanation in terms of the dynamics of the strongly interacting, nearly ideal, quark-gluon fluid undergoing rapid expansion. It was shown by Alver and collaborators~\cite{Alver:2010gr,Alver:2010dn} that the entire structure of the two particle correlation spectra in the $\Delta\eta-\Delta \phi$ plane is a consequence of the pressure driven response of the matter deposited in the collision to its initial geometry. Since the positions of the nucleons and the number of these that participate in the collision fluctuates from event to event, so too does the spatial shape of the hot and dense matter formed in the transverse plane of the collision. The eccentricities of these shapes in each event, represented by Fourier moments of their azimuthal distributions, are efficiently converted by hydrodynamics into momentum space azimuthal anisotropies. The two particle correlations that are measured can be expanded in the Fourier series,
\begin{eqnarray}
\frac{1}{N_{trig}N_{assoc.}} \frac{d^2N}{d\Delta \phi} \simeq 1+ 2\,V_1 \mathrm{cos}(\Delta \phi)+2\,V_2 \mathrm{cos}(2\Delta \phi)+... \,,
\label{eqn:vnfourier}
\end{eqnarray}
where the coefficients define the commonly measured ``flow'' or Fourier harmonics, $v_n \simeq V_n^{1/2}$~\cite{Chatrchyan:2013nka}.
A number of studies have shown that the fluctuation driven anisotropies tend to be universal such that $\varepsilon_n\{m\}$ are equal for $m\geq4$~\cite{Bilandzic:2010jr,Yan:2013laa,Bzdak:2013raa}. As conjectured by Alver et al., detailed hydrodynamic computations reveal that the flow harmonics are well-reproduced~\cite{Gale:2013da} with a very low value of the shear viscosity to entropy ratio close to a conjectured universal lower bound for strongly interacting fluids~\cite{Kovtun:2004de}. Moreover, these computations are seen to be sensitive to sub-nucleon degrees of freedom, in particular the gluon saturation scale $Q_s$ inside each of the colliding nuclei~\cite{Schenke:2012wb,Gale:2012rq,Niemi:2015qia}. 

For smaller collision systems such as peripheral nucleus-nucleus collisions, proton--proton ($pp$) or proton--heavy-ion ($pA$) collisions, the applicability of hydrodynamics is pushed to the extreme because the Knudsen numbers that characterize these fluid become large~\cite{Bzdak:2013zma,Niemi:2014wta}. While hydrodynamics may still be applicable, higher order contributions are not under control; this is a situation that bears strong analogy~\cite{Heller:2013fn} to perturbative QCD when the coupling constant becomes large. Nevertheless, strikingly, a near-side as well as an away-side ridge is observed in high multiplicity $pp$~\cite{Khachatryan:2010gv} and $pA$ collisions~\cite{CMS:2012qk}. Even more remarkably, multiparticle correlations of four or more particles have also been observed in $pp$~\cite{Khachatryan:2016txc} and $pA$ collisions~\cite{Khachatryan:2015waa} with similar ordering of the $v_2\{m\}$ as in $AA$ collisions. These have been observed across a large range of beam energies at RHIC, as well as different systems, including proton/deuteron/$^3$He--heavy-ion collisions~\cite{Belmont:2017lse}. 

Despite the above stated caveats, the experimental data drive us to ask whether the data in small systems can be described by hydrodynamics and what the limits of its applicability are. First studies performed for $pA$ collisions indicated that some of the features of the data are consistent with hydro computations~\cite{Bozek:2012gr}. However it soon became clear that the results are very sensitive to the choice of initial conditions. In the IP-Glasma model that includes subnucleon fluctuations~\cite{Schenke:2012wb,Schenke:2012fw} and provides a good description of the $AA$ data, the choice of an isotropic color charge distribution inside the proton does not describe the data on $pA$ collisions~\cite{Schenke:2014zha}. One possible interpretation of this result is that shape fluctuations of the proton's color charge density are important~\cite{Schenke:2014zha,Bjorken:2013boa}. Indeed, better agreement  with the $v_n\{2\}$ data in $pA$ collisions~\cite{Mantysaari:2017cni} is found by taking into account  fluctuations in the shapes of the color charge distributions within the proton. The data on incoherent exclusive vector meson production from HERA are best fit by models that take such fluctuations in account~\cite{Mantysaari:2016ykx}. 

This strong sensitivity to shape fluctuations suggests that the initial conditions and their evolution with time are especially important for the smaller systems~\cite{Greif:2017bnr}. Further, despite the near equality of the $v_n\{m\}$ coefficients for $m\geq 4$ as a function of the event multiplicity $N_{\rm charge}$, this result is not simply recovered in {\it ab initio} hydro models. For $pp$ collisions, a recent detailed study shows that while the two particle $v_n$ data are well reproduced, this is not true for the four particle cumulants~\cite{Zhao:2017rgg} which are imaginary for large $N_{\rm charge}$. In $pA$ collisions, there have been very limited studies of four-particle cumulants with Glauber model initial conditions~\cite{Bozek:2013uha,Kozlov:2014fqa}; indeed, it has been noted that in this case the $pA$ and $AA$ data cannot be described by the same set of parameters~\cite{Kozlov:2014hya}. 

It is then natural to see to what extent features of initial state correlations can explain the observed correlations and seemingly collective behavior in small systems. Detailed models for initial state correlations can be extracted from a high energy effective theory of QCD, the Color Glass Condensate (CGC)~\cite{Iancu:2003xm,Gelis:2010nm}. What these models have in common is that key elements of their dynamics are described by an emergent saturation scale $Q_{s,T}$ in the target. The interplay of this emergent scale with the other scales in the problem can generate distinct behaviour in different kinematic regions. For instance, multiparticle dynamics in the regime of soft momenta $p_T < Q_{s,T}$ is sensitive to the dynamics of several color domains of size $1/Q_{s,T}$; in contrast, for $p_T \geq Q_{s,T}$, the physics is sensitive to dynamics within a given domain. For large $p_T > Q_{s,T}$, the quasiparticle nature of color fields is relevant and the QCD dynamics can be represented in terms of Feynman diagrams, the so-called Glasma graphs~\cite{Dumitru:2008wn,Dusling:2009ar,Dumitru:2010iy,Dusling:2012iga,Dusling:2012cg,Dusling:2012wy,Dusling:2013qoz,Dusling:2015rja,Ozonder:2017moj}. In the kinematic regime of 
$p_T \gg Q_{s,T}$, they provide a good description of the ridge.  Note that for gluons, power corrections of order $Q_{s,P}/p_T$, where $Q_{s,P}$ is the saturation scale in the projectile, are important for understanding odd anisotropies such as $v_3$, which are smaller than $v_2$ in this kinematic domain.  

For $p_T < Q_{s,T}$, the physics of multiple color domains can only be described by color fields and not by quark and gluon quasiparticles. In this regime, gluon field strengths are as large as permissible in QCD; the dynamics is therefore captured by the classical Yang-Mills equations~\cite{McLerran:1993ka,McLerran:1993ni,Kovchegov:1996ty}. In general, if the sources of color charge have a high phase-space occupancy, as is the case for $AA$ collisions, the Yang-Mills equations have to be solved numerically~\cite{Krasnitz:1998ns,Krasnitz:2001qu,Krasnitz:2003jw,Lappi:2003bi,Lappi:2007ku}. This is implemented for instance in the IP-Glasma model we discussed earlier.  For $pA$ collisions, if the $p_T$ in the projectile is small compared to $Q_{s,P}$ (relevant for the $p_T$ integrated $v_n\{m\}$), one needs to solve the Yang-Mills equations also for the $pA$ case. This was done in \cite{Schenke:2015aqa} where it was shown that significant $v_{2,3}(p_T)$ of gluons is generated in $pA$ collisions. In \cite{Schenke:2016lrs}, an event generator was developed whereby gluons produced in the Yang-Mills evolution were fragmented using the Lund string fragmentation algorithm~\cite{Andersson:1983ia,Sjostrand:1982fn}. This CGC+PYTHIA framework, applied to high multiplicity $pp$ collisions generates the near-side ridge observed in experiments.  The $v_2(p_T)$ extracted for different hadrons displays the mass splitting seen in the data; this is also seen in the flavor dependence of $\langle p_T\rangle$. Applying this framework directly to study multiparticle collectivity is challenging because it is very numerically intensive. (The same problem bedevils event-by-event hydrodynamic computations of multiparticle correlations in $pA$ collisions.) Very recent developments suggest however a considerable simplification to numerical solutions of the Yang-Mills equations that enable efficient computations of multiparticle azimuthal anisotropies~\cite{McLerran:2016snu}; these shall not be discussed at length any further but will be discussed separately~\cite{Skokov,Mace-Skokov-Tribedy-RV}.

We will discuss here instead a very simple model of eikonal quarks from the projectile proton scattering off a dense, gluon saturated heavy-ion target~\cite{Dusling:2017dqg,Dusling:2017aot}.  We are able to qualitatively explain with this model several of the signatures presented previously as {\it prima facie} evidence of hydrodynamic collectivity. This first proof-of-principle computation demonstrating that multiparticle correlations in the initial state can reproduce the systematics of the data suggests that more stringent tests of hydrodynamic collectivity are necessary; collectivity, as commonly defined empirically, can have more than one origin. We will also discuss some of the shortcomings of this simple model and how these demand a better understanding of the spatial and momentum structure of multiparton correlations in hadron wavefunctions.

\section{Multi-particle correlations and collectivity}

Multiparticle correlations have become an important tool for extracting the collective properties of the matter created after a heavy-ion collision. From the $m$-particle ($m$ even) generalization of the Fourier expansion of the two particle distribution in~\Eq{eqn:vnfourier}, it is possible to define cumulants of this $m$-particle distribution as
\begin{eqnarray}
c_n\{m\}= \langle \langle e^{in(\phi_1+...+\phi_{m/2}-\phi_{{m+1}/2}-...-\phi_{m})} \rangle \rangle \,.
\end{eqnarray}
The double expectation value $\langle\langle \cdots\rangle\rangle$ represents only connected contributions (thereby subtracting off all combinations of lesser particle number correlations) as well as averaging with respect to single events and over all events~\cite{Chatrchyan:2013nka}.

It has been shown that an estimate of $v_n$ can be obtained from a measurement of multi-particle cumulants $c_n\{m\}$, which by design removes so called nonflow contributions~\cite{Borghini:2001vi}
\begin{eqnarray}
v_n\{2\}^2=c_n\{2\}, \, v_n\{4\}^4=-c_n\{4\}, \, v_n\{6\}^6=c_n\{6\}/4, ... \,.
\end{eqnarray}
In such collisions, the $n$th Fourier moment of the spatial eccentricity is determined by performing the azimuthal angle average over the positions of the nucleons in each nuclei, 
\begin{eqnarray}
\epsilon_n=\frac{1}{\langle r_\perp^n \rangle} \int d^2r_\perp e^{in\phi_r} r_\perp^n \frac{dN}{dy d^2r_\perp}\,.
\end{eqnarray}
A Gaussian model of eccentricity distributions~\cite{Bzdak:2013rya,Bzdak:2013raa} then gives 
\begin{eqnarray}
\epsilon_n\{2\} > \epsilon_n\{4\} = \epsilon_n\{6\} = \epsilon_n\{8\} = ... \,.
\end{eqnarray}
In a standard picture of a single collective fluid described by hydrodynamics, initial spatial anisotropies are transmitted by a strongly interacting fluid to  momentum space anisotropy, resulting in~\cite{Gardim:2011xv,Niemi:2012aj,Yan:2013laa} 
\begin{eqnarray}
v_n\{m\} \simeq c_n \epsilon_n\{m\} \,.
\end{eqnarray}
Hence, as mentioned previously, hydrodynamic collectivity in this case would be characterized by multiparticle $v_2$ moments satisfying 
\begin{eqnarray}
v_2\{2\} > v_2\{4\} = v_2\{6\} = v_2\{8\} =... \,.
\end{eqnarray}
Underlying all this is a picture of correlations where the mean and the variance of the eccentricity distribution dominate over higher moments; this is 
likely the case when the number of fluctuating sources is large~\cite{Blaizot:2014nia,Gronqvist:2016hym}. While this is superficially like the McLerran-Venugopalan (MV) model~\cite{McLerran:1993ka,McLerran:1993ni}, there are essential and interesting ways in which the two pictures diverge.

\section{Parton model for multiparticle correlations}

As noted, employing numerical Yang-Mills simulations to extract multiparticle azimuthal anisotropies is computationally daunting~\cite{Schenke:2015aqa}. Instead, one can first investigate in a simple parton model whether initial state correlations can reproduce the systematics of multiparticle collectivity. Such a model was considered previously for two-particle $v_n\{2\}$ anisotropies in~\cite{Lappi:2015vha,Lappi:2015vta} and we will discuss here its generalization to compute $v_n\{m\}$ for 
$m\geq 4$~\cite{Dusling:2017dqg,Dusling:2017aot}. Consider $m$ nearly collinear quarks in the projectile which multiple scatter off a nuclear target containing coherent color charge domains of size $\sim1/Q_{s,T}$. In a high energy eikonal approximation, the partons are color rotated by the field of the target by a lightlike Wilson line
\begin{eqnarray}
U(\mathbf{x}_\perp)= \mathcal{P}\text{exp}\left(-ig\int dz^+ A^-_a(z^+,\mathbf{x}_\perp)\tau^a\right) \,,
\label{eqn:wilson_line}
\end{eqnarray}
where the $\tau^a$ with $a=1,\cdots,8$ are Gell-Mann matrices in the fundamental $SU(3)$ color representation and $A^-$ is the classical gauge field corresponding to color sources in the target. 
The single particle distribution is given by~\cite{Dumitru:2002qt,Lappi:2015vha}
\begin{eqnarray}
\left \langle \frac{dN_q}{d^2\mathbf{p}}\right \rangle =\frac{1}{4\pi B_p} \int d^2b \int d^2r \int \frac{d^2k}{(2\pi)^2} ~ \left[ \frac{1}{\pi^2} e^{-|\mathbf{b}|^2/B_p} e^{-|\mathbf{k}|^2 B_p}\right] e^{i \mathbf{r}\cdot(\mathbf{p}-\mathbf{k})} \left\langle D\left(\mathbf{b}+\frac{\mathbf{r}}{2},\mathbf{b}-\frac{\mathbf{r}}{2}\right) \right\rangle , \, \, \, \, \, \, \,  \, \, \, 
\label{eqn:singlepartdist}
\end{eqnarray}
where the target dependence is expressed in terms of the ensemble average (over color sources, or equivalently $A^-$) of the dipole operator $D(x,y)=\frac{1}{N_c}\,{\rm Tr}(U^\dagger(x)U(y))$.  The term in the square bracket is the Wigner function modeling the projectile. In this simple model, one nonperturbative scale $B_p$ sets both the size of the projectile, as well as the intrinsic transverse momenta of  quarks in the projectile. For all the calculations discussed here, we will fix $B_p=4 ~\text{GeV}^{-2}$; this choice is determined by dipole model fits to HERA DIS data on exclusive vector meson production~\cite{Kowalski:2006hc,Rezaeian:2012ji}. 

Multiparticle distributions are obtained analogously by taking the expectation value over $m$ products of the single particle distributions in~\Eq{eqn:singlepartdist}~\cite{Dusling:2017dqg,Dusling:2017aot}:
\begin{eqnarray}
\label{eqn:multiplicity_kintegrated}
\left\langle\frac{d^{m} N}{d^2\mathbf{p_{1}}\cdots d^2\mathbf{p_{m}} } \right\rangle&=& \frac{1}{(4\pi^3 B_p)^m} {\prod_{i=1}^m} \int d^2\mathbf{b_i}  \int d^2\mathbf{r_i}~e^{-|\mathbf{b_i}|^2/B_p} e^{-|\mathbf{r_i}|^2/4B_p}  e^{i\mathbf{p_{i}}\cdot \mathbf{r_i}} \nonumber \\
&\times& \left< \prod_{j=1}^m D\left(\mathbf{b_j} + \frac{\mathbf{r_j}}{2},\mathbf{b_j} - \frac{\mathbf{r_j}}{2}\right) \right>\,.
\end{eqnarray} 
The form of the l.h.s. for m-particle distributions is quite general~\cite{Dumitru:2008wn,Gelis:2008ad,Dusling:2009ni}. The r.h.s appears natural in a hybrid 
``dilute-dense" framework; it is however oversimplified and there may be significant corrections in general~\cite{Gelis:2008ad,Kovchegov:2012nd,Kovner:2017vro,Kovner:2018vec}. We will return to this point later.

With this expression for the $m$-particle distribution, we can proceed to discuss the computation of the azimuthal angular cumulants and harmonics. 
We will first have to compute the expectation value of the $m$-dipole correlator in~\Eq{eqn:multiplicity_kintegrated}. In the MV model, this is known for one and two dipoles for arbitrary $N_c$~\cite{Kovner:2001vi,Blaizot:2004wu,Dominguez:2008aa,Fukushima:2007dy}. In~\cite{Dusling:2017aot}, we gave a general method to compute an arbitrary number of dipole expectation values in the MV model. This is accomplished by expanding the Wilson line given by~\Eq{eqn:wilson_line} around the last slice in the $z^+=\xi$ coordinate, $U(x_\perp)\simeq V(x_\perp)[1+igA^-_a(\xi,x)T^a+...]$, where $V(x_\perp)$ excludes this last slice. This expansion can be carried out systematically with products of the dipole operator, grouping terms that depend on the Wilson lines $U$ and $V$ separately, where $U$ can be considered as having one additional gluon exchanged with the target compared to $V$. Taking traces, the Fierz identity alters the color structure, resulting in contributions from higher-point correlation functions of Wilson lines. For the four-particle correlations of interest, four dipoles are considered, resulting in five distinct topologies, composed of all possible permutations of dipoles, quadrupoles, sextupoles, and octupoles containing eight total Wilson lines. This creates a system of correlation functions, which can be simultaneously considered as an $m!\times m!$ matrix describing all possible one gluon exchanges. Summing over an infinite number of exchanges, we are able to compute the full $m$-dipole correlation expectation value. For more details, examples, and an alternative diagrammatic derivation, see~\cite{Dusling:2017aot}. With this powerful machinery, it is now possible to calculate expectation values in a Gaussian MV-like model for traces of products of an arbitrary number of Wilson lines. For correlation functions involving two or four Wilson lines, it is possible to compute their expectation values analytically; for products of more than four Wilson lines, it is only feasible to do this numerically, or at large $N_c$~\cite{Fukushima:2017mko,Dominguez:2012ad}.

Using the definition of the $m$-particle spectra in~\Eq{eqn:multiplicity_kintegrated}, and the intermediate expression 
\begin{eqnarray}
\kappa_n\{m\} \equiv \int  \prod_{i=1}^m  d^2\mathbf{p_{i}}\, e^{i n(\phi_1+\cdots+\phi_{m/2-1}-\phi_{m/2}-\cdots -\phi_{m})} \left< \frac{d^m N}{\prod_{i=1}^m d^2\mathbf{p_{i}}} \right> \,,
\label{eqn:kappa}
\end{eqnarray}
we obtain the m-particle azimuthal anisotropy cumulants: 
\begin{eqnarray}
c_n\{2\}&=&\langle  e^{in(\phi_1-\phi_2)} \rangle \label{eqn:cn2} \equiv \frac{\kappa_n\{2\}}{\kappa_0\{2\}} \\
c_n\{4\}&=&\langle  e^{in(\phi_1+\phi_2-\phi_3-\phi_4)} \rangle - 2 \langle  e^{in(\phi_1-\phi_2)}\rangle^2 \equiv \frac{\kappa_n\{4\}}{\kappa_0\{4\}} - 2 \left( \frac{\kappa_n\{4\}}{\kappa_0\{4\}}\right)^2 \,.
\label{eqn:cn4}
\label{eqn:intrinsic-cumulants}
\end{eqnarray}
Higher point cumulants are defined analogously. In the spirit of what is done experimentally for the measured hadrons, we integrate over the momenta of the scattered quarks from zero to $p_\perp^{\text{max}}$ for the $p_T$ integrated moments. We are also interested in calculating cumulants differential in the momenta of one of the quarks; in this case, $m-1$ momentum integrals in \Eq{eqn:kappa} are considered.

\section{Results}
\begin{figure}
\begin{subfigure}[t]{0.48\textwidth}
\includegraphics[width=\textwidth]{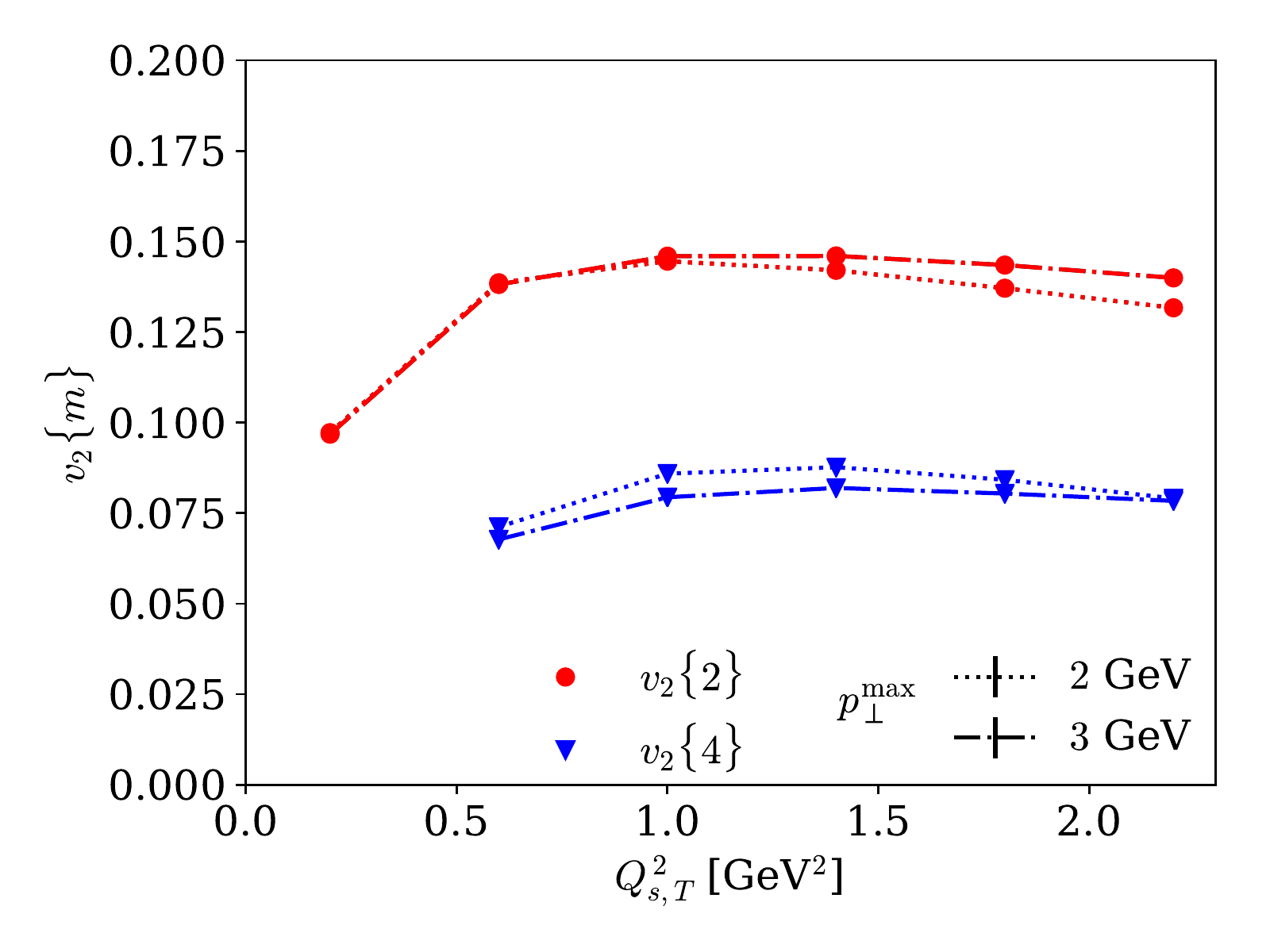}
\end{subfigure}
\begin{subfigure}[t]{0.48\textwidth}
\includegraphics[width=\textwidth]{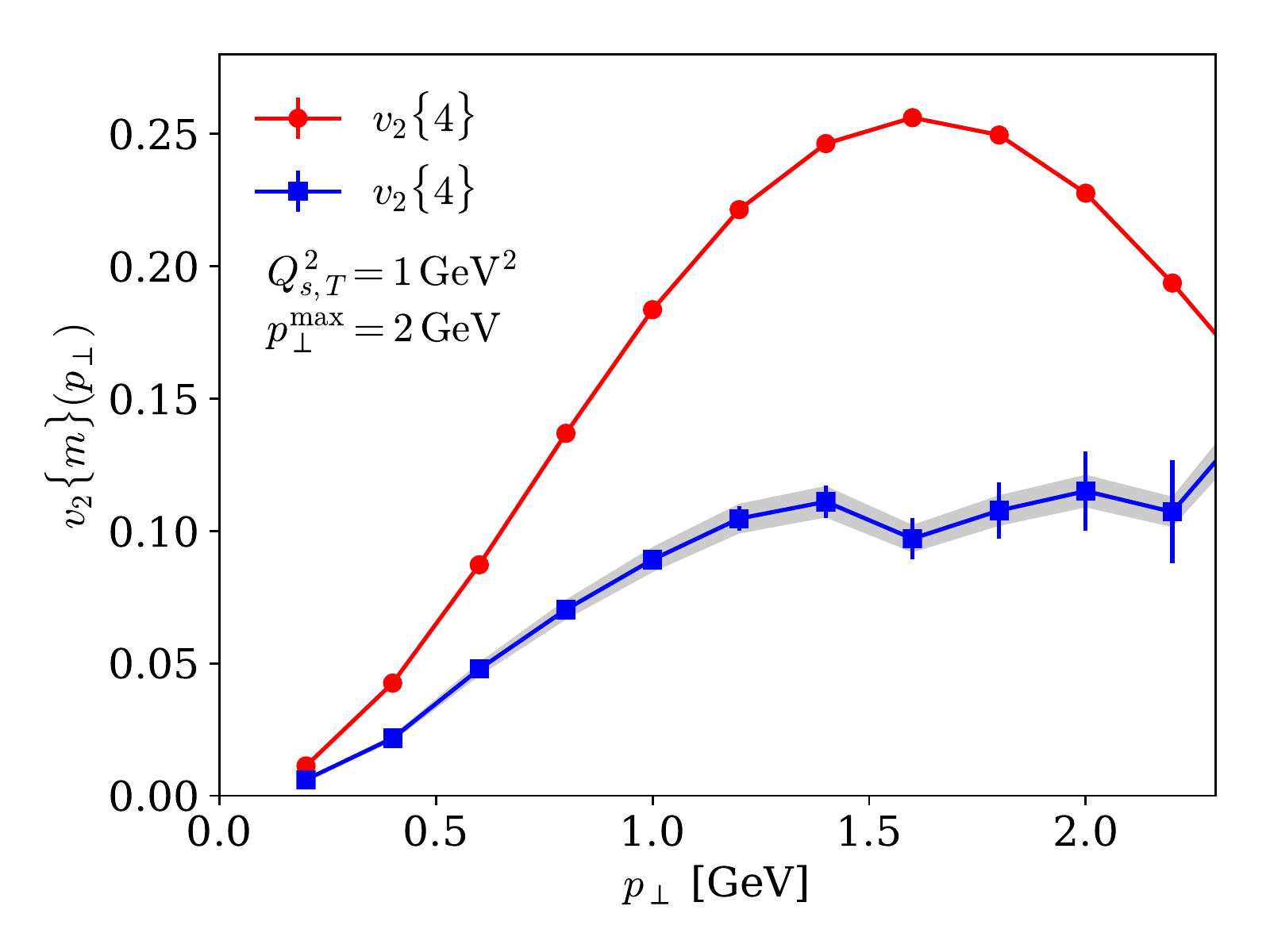}
\end{subfigure}
\caption{The Fourier harmonics $v_2\{2\}$ and $v_2\{4\}$. Left: Integrated of momentum as a function of $Q_{s,T}$. Right: As a function of $p_\perp$ for a given $Q_{s,T}$.}
\label{fig:v2}
\end{figure}

\begin{figure}
\begin{subfigure}[t]{0.48\textwidth}
\includegraphics[width=\textwidth]{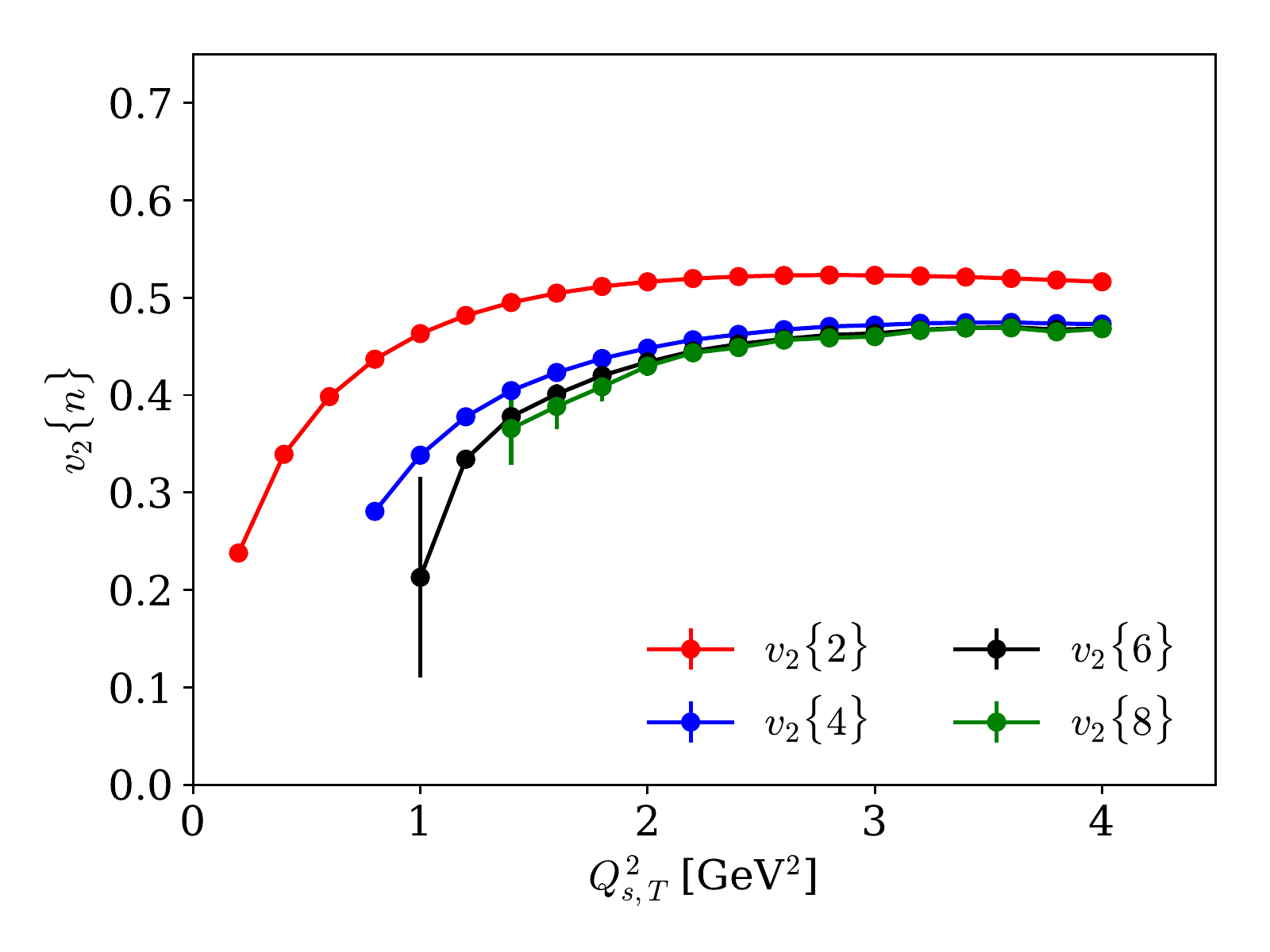}
\end{subfigure}
\begin{subfigure}[t]{0.48\textwidth}
\includegraphics[width=\textwidth]{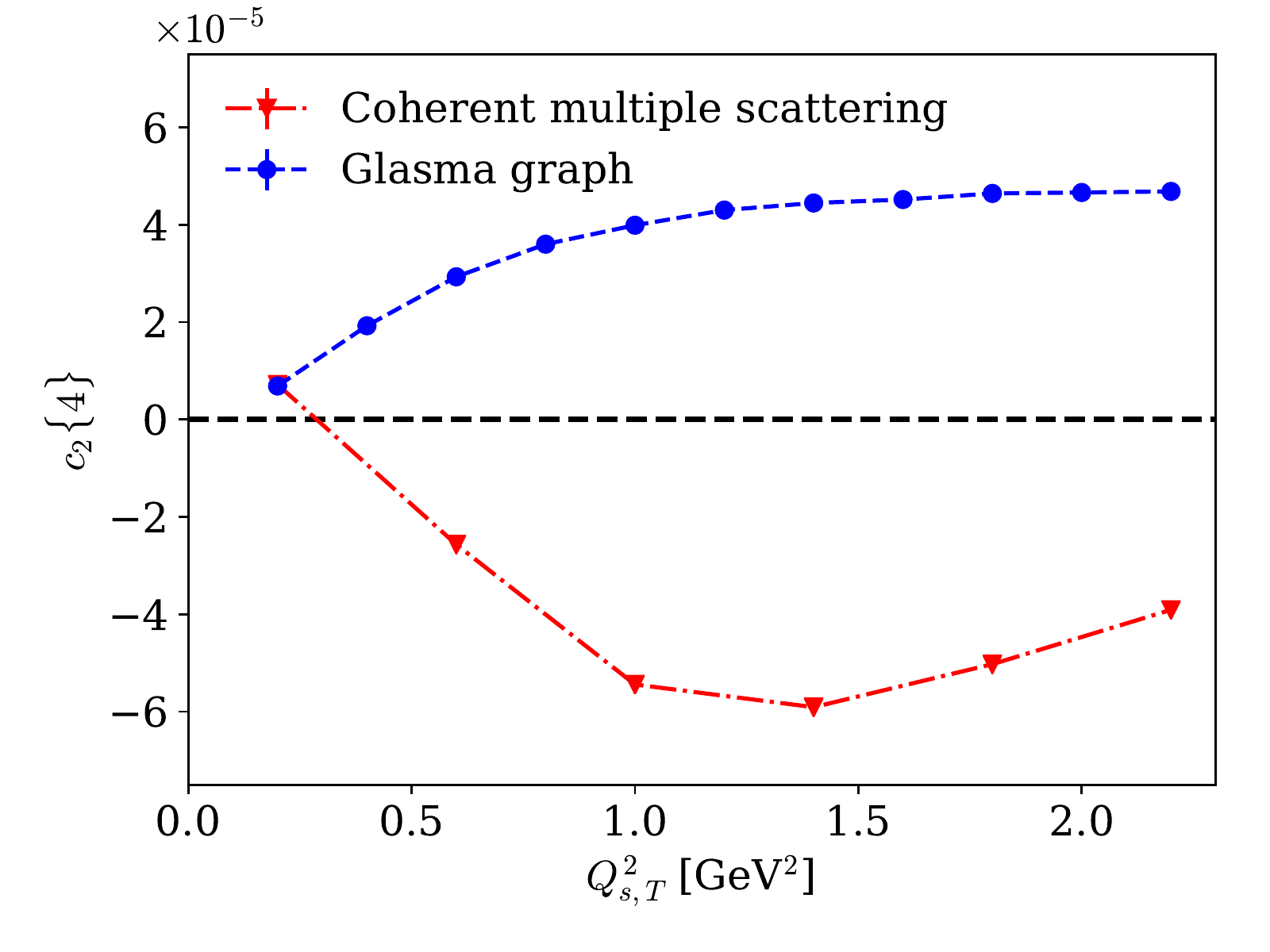}
\end{subfigure}
\caption{Left: Multi-particle second Fourier harmonics $v_2\{m\}$ in an Abelian version of our model. Right: The four-particle second cumulant $c_2\{4\}$ in the glasma graph approximation to our coherent multiple scattering model in comparison to the full model.}
\label{fig:v2other}
\end{figure}

\begin{figure}
\centering
\includegraphics[width=0.52\textwidth]{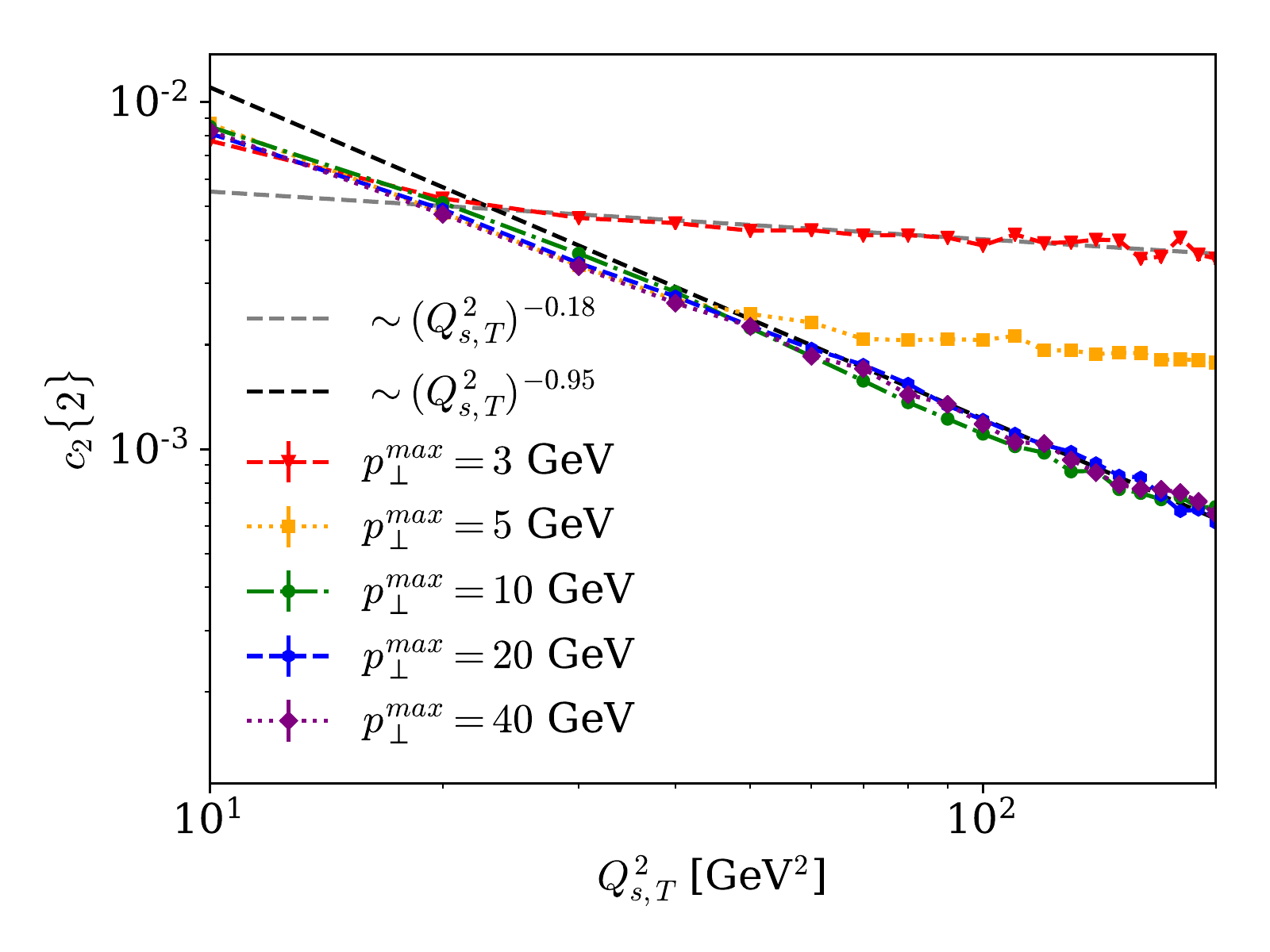}
\caption{Two particle cumulant $c_2\{2\}$ for various $p_\perp^{max}$ as a function of large values of $Q_{s,T}^2$. A fixed value of $\Bp=4~\text{GeV}^{-2}$ is used. For $\pmax\leq Q_{s,T}$, we see a weak dependence on $Q_{s,T}^2\Bp$. For larger $p_\perp^{max}$, we see a falloff in the value of the cumulant that scales approximately with the number of color domains as $\sim1/{Q_{s,T}^2\Bp}$.}
\label{fig:independent-cluster}
\end{figure}

We begin by studying the momentum-integrated anisotropy coefficients as a function of $Q_{s,T}$. This is shown in the left panel of \Fig{fig:v2}. In our model, the observed correlations are relatively insensitive to this maximum integrated momentum, $p_\perp^{\text{max}}$. There is a clear ordering in the two- versus four-particle $v_2$, as observed in the experimental data~\cite{Khachatryan:2015waa}. While by definition $v_2\{2\}$ is real for all values of $Q_{s,T}$, the appearance of a real $v_2\{4\}=(-c_2\{4\})^{1/4}$ corresponds to a negative $c_2\{4\}$. This negative $c_2\{4\}$ occurs at a finite value of $Q_{s,T}$, a trend that is qualitatively similar to the change from positive to negative values observed experimentally for $c_2\{4\}$ as a function of increasing multiplicity~\cite{Aaboud:2017blb}. While $Q_{s,T}$ is not directly representative of the multiplicity, the product $B_p Q_{x,T}^2$ corresponds to the number of domains that quarks from the projectile interact with. The saturation scale $Q_{s,T}$ grows with decreasing values of $x$ or with increasing center of mass energy.  The insensitivity of the cumulants to varying $Q_{s,T}^2$ is therefore consistent with their weak energy dependence observed in experiments~\cite{Khachatryan:2015waa,Belmont:2017lse}.  The transverse momentum dependent Fourier harmonics, shown in the right panel of \Fig{fig:v2}, are also similar in magnitude and shape to that observed experimentally~\cite{Chatrchyan:2013nka}, with a peak between $\sim 1-2~\text{GeV}$. Their magnitudes are typically larger than the data; these will be lowered and smeared by the fragmentation of the partons. 

To study whether the model displays the striking convergence of $v_2\{m\}$ for increasing values of $m$ seen in the data, we consider a $U(1)$ Abelian version of our model. Instead of computing traces of path ordered $SU(3)$ matrices, we have a much simpler product of exponentials enabling rapid computations for large values of $m$. The result is shown in the left panel of \Fig{fig:v2other}. We first note that the Abelian model qualitatively has the same structure for the two- and four-particle $v_2$ as the non-Abelian model results shown in \Fig{fig:v2}. However most strikingly, the six- and eight-particle $v_2$'s converge to the value for $v_2\{4\}$, mirroring what is seen in experiments~\cite{Khachatryan:2015waa}. When this convergence was first seen in experiment, it was seen by some to be definitive evidence of hydrodynamic collectivity. \Fig{fig:v2other} provides an extremely simple counter example that this pattern, in of itself, is not conclusive. We also show in \cite{Dusling:2017dqg,Dusling:2017aot} that the systematics of so-called symmetric cumulants, likewise seen as a consequence of hydodynamic response to geometry, are reproduced in our simple model. 

It is also instructive to compare our model to the glasma graph approximation. (For more details on this comparison, see the Appendix of~\cite{Dusling:2017aot}.) As noted previously, the glasma graph approximation is valid for $p_T > Q_{s,T}$. On the other hand, multiple scattering (higher twist) contributions are parametrically of order $Q_{s,T}/p_T$ and are therefore large for $p_T < Q_{s,T}$. The right panel of \Fig{fig:v2other} shows that there is a clear distinction between the two limits. The glasma graphs do not produce a real $v_2\{4\}$ in marked contrast to the limit of coherent multiple scattering. The latter is therefore a key feature of our model; it suppresses the large higher Glasma graph cumulants (the distribution of which is close to that of a Bose-distribution!~\cite{Gelis:2009wh}) relative to the lower cumulants, thereby generating positive $v_2\{m\}$ for $m\geq 4$. 

In our model, $\Bp$ represents the transverse overlap area of the projectile with the target and $1/Q_{s,T}^2$ sets the scale for the size of color domains in the target. Hence the dimensionless product $Q_{s,T}^2\Bp$ represents the number of color domains in the target that interact with the projectile constituents. 
Independent cluster models suggest that the number of domains should fall as $1/(Q_{s,T}^2\Bp)$.  However, as shown in \cite{Dusling:2017aot}, this is not seen for the integrated cumulants.  The reason is that there is another scale $\pmax$ controlling the maximal momentum kick from the target to the probe; its inverse is the smallest distance that the probe resolves in the target. There are therefore two dimensionless combinations, $Q_{s,T}^2 \Bp$ and $Q_{s,T}^2/({\pmax})^2$. The dependence of our results on the number of color domains depends on what $Q_{s,T}^2/({\pmax})^2$ is. 

For $({\pmax})^2 \gtrsim Q_{s,T}^2$,  the probe resolves an area within individual domains;  we therefore expect correlation strengths to fall approximately as $1/(Q_{s,T}^2\Bp)$.  \Fig{fig:independent-cluster} shows that $c_2\{2\}$ (for $\pmax=10, 20, 40$ GeV) satisfies the scaling form $(Q_{s,T}^2\Bp)^{-0.95}$ at large $Q_{s,T}^2$.  On the other hand, for $({\pmax})^2 \leq Q_{s,T}^2$, the probe only resolves transverse sizes larger than the typical domain size. For these smaller values of $\pmax$, increasing $Q_{s,T}^2\Bp$, and therefore the number of color domains, does not change the signal since the probe cannot resolve the change in the number of color domains. Indeed, \Fig{fig:independent-cluster} shows that for $\pmax=3,5$ GeV we see a rather modest falloff, $c_2\{2\} \sim (Q_{s,T}^2\Bp)^{-0.18}$, demonstrating that azimuthal cumulants are weakly dependent on the number of clusters~\cite{Basar:2013hea}. 

There is an oft repeated mantra that because initial state collectivity occurs on the scale $1/Q_{s,T}$, and the sizes of the transverse overlap area can be much larger, that correlations should die off when the latter increases. This intuition is based on the independent cluster model. As noted, our result in \Fig{fig:independent-cluster} shows that this is not the case when $\pmax$ is smaller than $Q_{s,T}$, with the result being only weakly dependent on the number of clusters. This is because the $p_T$ kick to the projectile is a collective effect of multiple domains in the target. 

When partons in the projectile are separated at distances greater than a fermi, as in a deuteron ($dA$) or Helium-3 ($^3He-A$) nucleus, they can still experience collective kicks from the target;  these are however color singlet exchanges and one expects the kicks to be significantly weaker. For minimum bias configurations of nucleons in such light-ion projectiles, even a collective effect from a common hydrodynamic response will be weak  because Hanbury-Brown--Twiss radii for $pA$ collisions indicate that the matter produced freezes out on distance scales of a fermi~\cite{Adam:2015pya}. However the ridge effect in such systems is most prominent for rare high multiplicity $dA$ and $^3He-A$ collisions, where the nucleons in the projectile are more closely packed than in typical events. Indeed, the multiplicity in such collisions is more strongly correlated with the saturation scale in the projectile than that in the target. Proper modeling of such effects requires a more sophisticated treatment of parton distributions in the projectile than is feasible in our simple model. This can however be achieved in the full Yang-Mills CGC framework and results from these simulations will soon be available.

\section{Concluding remarks}
%Discuss rapidity dependence

We  have shown in our simple initial state model that we are able to qualitatively reproduce many of the features associated with collectivity in small systems. However the observed correlations are long range in rapidity and we have neglected any discussion of rapidity in our model.  In the ``hybrid" framework,  valid in the forward region, we should consider quarks with relatively large $x_q$ (usually taken to be $x_q\geq 0.01$). These large $x_q$ quarks are most naturally taken to be valence quarks, whereas long-range correlations necessarily probe smaller $x_q$ quarks. This does not preclude long-range correlations in our model. In order to include the quark rapidity dependence, we consider the single quark distribution in~\Eq{eqn:singlepartdist} and convolve it with the quark distribution function $f(x_q)$  to write (in terms of the rapidity $y$ and quark transverse momentum $p_\perp$)~\cite{Dusling:2017aot}
\begin{equation}
\frac{dN^{pA\to q+X}}{dy d^2{\bf p}_\perp}=x_q f(x_q) \frac{dN^{qA\to q+X}}{d^2{\bf p}_\perp}\,; \qquad
x_q=\frac{p_\perp}{\sqrt{s}}e^y\,.
\end{equation}
This is straightforwardly extended to two and multiparticle distributions. 
\begin{figure}
\center
\includegraphics[width=\textwidth]{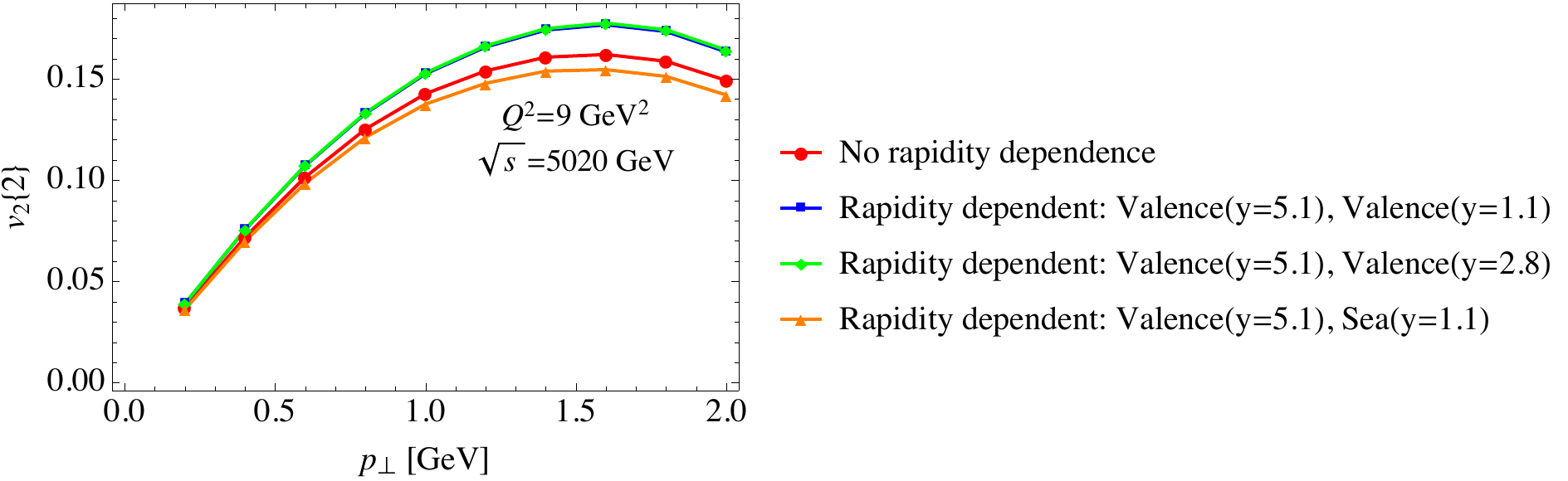}
\caption{Two-particle second Fourier harmonic $v_2\{2\}$ as a function of the momentum of one of the quarks both with and without rapidity dependence introduced via convolution with quark distributions.}
\label{fig:v22rap}
\end{figure}
For our results, we will employ the quark distribution functions of the NNPDF collaboration~\cite{Ball:2014uwa}.
While its straightforward to  possible include the fragmentation of the final state quarks into hadrons, this will not qualitatively change the rapidity dependence. 

Our computation for the rapidity dependence are shown in~\Fig{fig:v22rap}. We first consider a valence quark at $y=5.1$; at the top LHC energy of $\sqrt{s}=5.02~\text{TeV}$, this corresponds to an $x_q=0.1$. Then considering a separation in rapidity of 4 units, which corresponds to a second quark with $x_q\simeq 0.002$, we  clearly see that there is no quantitative difference with our previous result that neglected any rapidity dependence. Since it is unlikely a valence quark at this small $x$ can be found in the proton, we can also consider a valence quark with larger $x_q\simeq 0.01$, corresponding to $y=2.8$. There is no difference in the observed correlations for the two rapidity gaps considered. It is also interesting to consider correlations between valence and sea quarks. For four-particle correlations, if one does not consider correlations with and amongst gluons, this is the most likely scenario. \Fig{fig:v22rap} shows that all the differences are only quantitative, not qualitative. This is because when 
one takes the ratio of distributions to compute $c_2\{2\}$ (see~\Eq{eqn:cn2}), the rapidity dependence in our model all but disappears. 

A criticism that could be leveled at our model is that we consider quarks alone and not gluons in the projectile. At high energies,  gluons dominant hadron wavefunctions. However, we expect the underlying physics to remain qualitatively the same as previously. The azimuthal correlations result from time dilated partons being color rotated by a common strongly localized colored ``shock wave" in the nucleus and this is no less true for gluons than it is for quarks. These gluons will produce correlations which are inherently long-range in rapidity~\cite{Dusling:2009ni}; the quibble that they do not produce odd harmonics is removed as soon as one steps away from the strict dilute-dense limit~\cite{Schenke:2015aqa,McLerran:2016snu}. 

A more serious criticism is that the simple factorization in~\Eq{eqn:multiplicity_kintegrated} may not hold due to quantum interference effects~\cite{Kovner:2018vec,Kovner:2017vro}. While such effects are certainly there, there are open issues remaining whether they can be expressed as the generalized transverse momentum dependent (G-TMD) expressions as suggested in \cite{Kovner:2018vec,Kovner:2017vro} and what their impact on the results are. 
At any rate, these effects should be contained in the full Yang-Mills computation alluded to previously~\cite{Skokov,Mace-Skokov-Tribedy-RV}. Preliminary computations in the Yang-Mills framework suggest that the results are similar to those discussed here. The results of these computations will be discussed separately. 

Finally, one may ask whether the boost invariant Yang-Mills formalism is sufficient. If plasma instabilities occur rapidly on the time scales of the collision~\cite{Romatschke:2005pm,Romatschke:2006nk}, one may need to revert to the full 3+1 dimensional formalism~\cite{Berges:2013eia,Berges:2013fga}, which matches smoothly on to the bottom-up thermalization scenario~\cite{Baier:2000sb}. In this case, final state scattering may play a role by modifying the magnitude of the initially generated azimuthal anisotropy. Such a scenario has been modeled recently with promising results~\cite{Greif:2017bnr} and therefore demands closer investigation. 
\section{Acknowledgments}
This material is based on work supported by the U.S. Department of Energy, Office of Science, Office of Nuclear Physics, under Contracts No. DE-SC0012704 (M.M. and R.V.) and No. DE-FG02-88ER40388 (M.M.), and within the framework of the Beam Energy Scan Theory (BEST) and TMD Theory Topical Collaborations. This research used resources of the National Energy Research Scientific Computing Center, a DOE Office of Science User Facility supported by the Office of Science of the U.S. Department of Energy under Contract No. DE-AC02-05CH11231 and the LIRED computing system at the Institute for Advanced Computational Science at Stony Brook University.


\begin{thebibliography}{99}
  
  %\cite{Adare:2008ae}
\bibitem{Adare:2008ae} 
  A.~Adare {\it et al.} [PHENIX Collaboration],
  %``Dihadron azimuthal correlations in Au+Au collisions at s(NN)**(1/2) = 200-GeV,''
  Phys.\ Rev.\ C {\bf 78}, 014901 (2008)
  doi:10.1103/PhysRevC.78.014901
  [arXiv:0801.4545 [nucl-ex]].
  %%CITATION = doi:10.1103/PhysRevC.78.014901;%%
  %377 citations counted in INSPIRE as of 26 Jan 2018
  
  %\cite{Alver:2009id}
\bibitem{Alver:2009id} 
  B.~Alver {\it et al.} [PHOBOS Collaboration],
  %``High transverse momentum triggered correlations over a large pseudorapidity acceptance in Au+Au collisions at s(NN)**1/2 = 200 GeV,''
  Phys.\ Rev.\ Lett.\  {\bf 104}, 062301 (2010)
  doi:10.1103/PhysRevLett.104.062301
  [arXiv:0903.2811 [nucl-ex]].
  %%CITATION = doi:10.1103/PhysRevLett.104.062301;%%
  %221 citations counted in INSPIRE as of 26 Jan 2018
  
%\cite{Abelev:2009af}
\bibitem{Abelev:2009af} 
  B.~I.~Abelev {\it et al.} [STAR Collaboration],
  %``Long range rapidity correlations and jet production in high energy nuclear collisions,''
  Phys.\ Rev.\ C {\bf 80}, 064912 (2009)
  doi:10.1103/PhysRevC.80.064912
  [arXiv:0909.0191 [nucl-ex]].
  %%CITATION = doi:10.1103/PhysRevC.80.064912;%%
  %273 citations counted in INSPIRE as of 27 Dec 2017
  
  %\cite{Alver:2010gr}
\bibitem{Alver:2010gr}
  B.~Alver and G.~Roland,
  %``Collision geometry fluctuations and triangular flow in heavy-ion collisions,''
  Phys.\ Rev.\ C {\bf 81} (2010) 054905
   Erratum: [Phys.\ Rev.\ C {\bf 82} (2010) 039903]
  doi:10.1103/PhysRevC.82.039903, 10.1103/PhysRevC.81.054905
  [arXiv:1003.0194 [nucl-th]].
  %%CITATION = doi:10.1103/PhysRevC.82.039903, 10.1103/PhysRevC.81.054905;%%
  %606 citations counted in INSPIRE as of 27 Dec 2017
  
  %\cite{Alver:2010dn}
\bibitem{Alver:2010dn} 
  B.~H.~Alver, C.~Gombeaud, M.~Luzum and J.~Y.~Ollitrault,
  %``Triangular flow in hydrodynamics and transport theory,''
  Phys.\ Rev.\ C {\bf 82}, 034913 (2010)
  doi:10.1103/PhysRevC.82.034913
  [arXiv:1007.5469 [nucl-th]].
  %%CITATION = doi:10.1103/PhysRevC.82.034913;%%
  %280 citations counted in INSPIRE as of 27 Dec 2017
  
      
    %\cite{Chatrchyan:2013nka}
\bibitem{Chatrchyan:2013nka} 
  S.~Chatrchyan {\it et al.} [CMS Collaboration],
  %``Multiplicity and transverse momentum dependence of two- and four-particle correlations in pPb and PbPb collisions,''
  Phys.\ Lett.\ B {\bf 724}, 213 (2013)
  doi:10.1016/j.physletb.2013.06.028
  [arXiv:1305.0609 [nucl-ex]].
  %%CITATION = doi:10.1016/j.physletb.2013.06.028;%%
  %238 citations counted in INSPIRE as of 20 Jun 2016
  
  %\cite{Bilandzic:2010jr}
\bibitem{Bilandzic:2010jr} 
  A.~Bilandzic, R.~Snellings and S.~Voloshin,
  %``Flow analysis with cumulants: Direct calculations,''
  Phys.\ Rev.\ C {\bf 83}, 044913 (2011)
  doi:10.1103/PhysRevC.83.044913
  [arXiv:1010.0233 [nucl-ex]].
  %%CITATION = doi:10.1103/PhysRevC.83.044913;%%
  %228 citations counted in INSPIRE as of 26 Jan 2018

    
    %\cite{Yan:2013laa}
\bibitem{Yan:2013laa} 
  L.~Yan and J.~Y.~Ollitrault,
  %``Universal fluctuation-driven eccentricities in proton-proton, proton-nucleus and nucleus-nucleus collisions,''
  Phys.\ Rev.\ Lett.\  {\bf 112}, 082301 (2014)
  doi:10.1103/PhysRevLett.112.082301
  [arXiv:1312.6555 [nucl-th]].
  %%CITATION = doi:10.1103/PhysRevLett.112.082301;%%
  %69 citations counted in INSPIRE as of 27 Dec 2017
  
  %\cite{Bzdak:2013raa}
\bibitem{Bzdak:2013raa} 
  A.~Bzdak and V.~Skokov,
  %``Multi-particle eccentricities in collisions dominated by fluctuations,''
  Nucl.\ Phys.\ A {\bf 943}, 1 (2015)
  doi:10.1016/j.nuclphysa.2015.08.001
  [arXiv:1312.7349 [hep-ph]].
  %%CITATION = doi:10.1016/j.nuclphysa.2015.08.001;%%
  %15 citations counted in INSPIRE as of 27 Dec 2017
  
  %\cite{Gale:2013da}
\bibitem{Gale:2013da} 
  C.~Gale, S.~Jeon and B.~Schenke,
  %``Hydrodynamic Modeling of Heavy-Ion Collisions,''
  Int.\ J.\ Mod.\ Phys.\ A {\bf 28}, 1340011 (2013)
  doi:10.1142/S0217751X13400113
  [arXiv:1301.5893 [nucl-th]].
  %%CITATION = doi:10.1142/S0217751X13400113;%%
  %313 citations counted in INSPIRE as of 26 Jan 2018

%\cite{Kovtun:2004de}
\bibitem{Kovtun:2004de} 
  P.~Kovtun, D.~T.~Son and A.~O.~Starinets,
  %``Viscosity in strongly interacting quantum field theories from black hole physics,''
  Phys.\ Rev.\ Lett.\  {\bf 94}, 111601 (2005)
  doi:10.1103/PhysRevLett.94.111601
  [hep-th/0405231].
  %%CITATION = doi:10.1103/PhysRevLett.94.111601;%%
  %1943 citations counted in INSPIRE as of 26 Jan 2018
  
  %\cite{Schenke:2012wb}
\bibitem{Schenke:2012wb} 
  B.~Schenke, P.~Tribedy and R.~Venugopalan,
  %``Fluctuating Glasma initial conditions and flow in heavy ion collisions,''
  Phys.\ Rev.\ Lett.\  {\bf 108}, 252301 (2012)
  doi:10.1103/PhysRevLett.108.252301
  [arXiv:1202.6646 [nucl-th]].
  %%CITATION = doi:10.1103/PhysRevLett.108.252301;%%
  %278 citations counted in INSPIRE as of 26 Jan 2018
  
     
   %\cite{Gale:2012rq}
\bibitem{Gale:2012rq} 
  C.~Gale, S.~Jeon, B.~Schenke, P.~Tribedy and R.~Venugopalan,
  %``Event-by-event anisotropic flow in heavy-ion collisions from combined Yang-Mills and viscous fluid dynamics,''
  Phys.\ Rev.\ Lett.\  {\bf 110}, no. 1, 012302 (2013)
  doi:10.1103/PhysRevLett.110.012302
  [arXiv:1209.6330 [nucl-th]].
  %%CITATION = doi:10.1103/PhysRevLett.110.012302;%%
  %354 citations counted in INSPIRE as of 09 Jan 2018

  %\cite{Niemi:2015qia}
\bibitem{Niemi:2015qia} 
  H.~Niemi, K.~J.~Eskola and R.~Paatelainen,
  %``Event-by-event fluctuations in a perturbative QCD + saturation + hydrodynamics model: Determining QCD matter shear viscosity in ultrarelativistic heavy-ion collisions,''
  Phys.\ Rev.\ C {\bf 93}, no. 2, 024907 (2016)
  doi:10.1103/PhysRevC.93.024907
  [arXiv:1505.02677 [hep-ph]].
  %%CITATION = doi:10.1103/PhysRevC.93.024907;%%
  %115 citations counted in INSPIRE as of 26 Jan 2018
  
  %\cite{Bzdak:2013zma}
\bibitem{Bzdak:2013zma} 
  A.~Bzdak, B.~Schenke, P.~Tribedy and R.~Venugopalan,
  %``Initial state geometry and the role of hydrodynamics in proton-proton, proton-nucleus and deuteron-nucleus collisions,''
  Phys.\ Rev.\ C {\bf 87}, no. 6, 064906 (2013)
  doi:10.1103/PhysRevC.87.064906
  [arXiv:1304.3403 [nucl-th]].
  %%CITATION = doi:10.1103/PhysRevC.87.064906;%%
  %193 citations counted in INSPIRE as of 08 Jan 2018
  
  %\cite{Niemi:2014wta}
\bibitem{Niemi:2014wta} 
  H.~Niemi and G.~S.~Denicol,
  %``How large is the Knudsen number reached in fluid dynamical simulations of ultrarelativistic heavy ion collisions?,''
  arXiv:1404.7327 [nucl-th].
  %%CITATION = ARXIV:1404.7327;%%
  %42 citations counted in INSPIRE as of 08 Jan 2018
  
    %\cite{Heller:2013fn}
\bibitem{Heller:2013fn} 
  M.~P.~Heller, R.~A.~Janik and P.~Witaszczyk,
  %``Hydrodynamic Gradient Expansion in Gauge Theory Plasmas,''
  Phys.\ Rev.\ Lett.\  {\bf 110}, no. 21, 211602 (2013)
  doi:10.1103/PhysRevLett.110.211602
  [arXiv:1302.0697 [hep-th]].
  %%CITATION = doi:10.1103/PhysRevLett.110.211602;%%
  %84 citations counted in INSPIRE as of 26 Jan 2018
  
  %\cite{Khachatryan:2010gv}
\bibitem{Khachatryan:2010gv} 
  V.~Khachatryan {\it et al.} [CMS Collaboration],
  %``Observation of Long-Range Near-Side Angular Correlations in Proton-Proton Collisions at the LHC,''
  JHEP {\bf 1009}, 091 (2010)
  doi:10.1007/JHEP09(2010)091
  [arXiv:1009.4122 [hep-ex]].
  %%CITATION = doi:10.1007/JHEP09(2010)091;%%
  %629 citations counted in INSPIRE as of 27 Dec 2017
  
  %\cite{CMS:2012qk}
\bibitem{CMS:2012qk} 
  S.~Chatrchyan {\it et al.} [CMS Collaboration],
  %``Observation of long-range near-side angular correlations in proton-lead collisions at the LHC,''
  Phys.\ Lett.\ B {\bf 718}, 795 (2013)
  doi:10.1016/j.physletb.2012.11.025
  [arXiv:1210.5482 [nucl-ex]].
  %%CITATION = doi:10.1016/j.physletb.2012.11.025;%%
  %480 citations counted in INSPIRE as of 09 Jan 2018
  
%\cite{Khachatryan:2016txc}
\bibitem{Khachatryan:2016txc} 
  V.~Khachatryan {\it et al.} [CMS Collaboration],
  %``Evidence for collectivity in pp collisions at the LHC,''
  Phys.\ Lett.\ B {\bf 765}, 193 (2017)
  doi:10.1016/j.physletb.2016.12.009
  [arXiv:1606.06198 [nucl-ex]].
  %%CITATION = doi:10.1016/j.physletb.2016.12.009;%%
  %100 citations counted in INSPIRE as of 09 Jan 2018
  
  %\cite{Khachatryan:2015waa}
\bibitem{Khachatryan:2015waa} 
  V.~Khachatryan {\it et al.} [CMS Collaboration],
  %``Evidence for Collective Multiparticle Correlations in p-Pb Collisions,''
  Phys.\ Rev.\ Lett.\  {\bf 115}, no. 1, 012301 (2015)
  doi:10.1103/PhysRevLett.115.012301
  [arXiv:1502.05382 [nucl-ex]].
  %%CITATION = doi:10.1103/PhysRevLett.115.012301;%%
  %84 citations counted in INSPIRE as of 01 May 2017
  
  %\cite{Belmont:2017lse}
\bibitem{Belmont:2017lse} 
  R.~Belmont [PHENIX Collaboration],
  %``PHENIX results on multiparticle correlations in small systems,''
  Nucl.\ Phys.\ A {\bf 967}, 341 (2017)
  doi:10.1016/j.nuclphysa.2017.04.044
  [arXiv:1704.04570 [nucl-ex]].
  %%CITATION = doi:10.1016/j.nuclphysa.2017.04.044;%%
  %4 citations counted in INSPIRE as of 14 Dec 2017
  
  %\cite{Bozek:2012gr}
\bibitem{Bozek:2012gr} 
  P.~Bozek and W.~Broniowski,
  %``Correlations from hydrodynamic flow in p-Pb collisions,''
  Phys.\ Lett.\ B {\bf 718}, 1557 (2013)
  doi:10.1016/j.physletb.2012.12.051
  [arXiv:1211.0845 [nucl-th]].
  %%CITATION = doi:10.1016/j.physletb.2012.12.051;%%
  %183 citations counted in INSPIRE as of 09 Jan 2018
  
  %\cite{Schenke:2012fw}
\bibitem{Schenke:2012fw} 
  B.~Schenke, P.~Tribedy and R.~Venugopalan,
  %``Event-by-event gluon multiplicity, energy density, and eccentricities in ultrarelativistic heavy-ion collisions,''
  Phys.\ Rev.\ C {\bf 86}, 034908 (2012)
  doi:10.1103/PhysRevC.86.034908
  [arXiv:1206.6805 [hep-ph]].
  %%CITATION = doi:10.1103/PhysRevC.86.034908;%%
  %185 citations counted in INSPIRE as of 26 Jan 2018
 

%\cite{Schenke:2014zha}
\bibitem{Schenke:2014zha} 
  B.~Schenke and R.~Venugopalan,
  %``Eccentric protons? Sensitivity of flow to system size and shape in p+p, p+Pb and Pb+Pb collisions,''
  Phys.\ Rev.\ Lett.\  {\bf 113}, 102301 (2014)
  doi:10.1103/PhysRevLett.113.102301
  [arXiv:1405.3605 [nucl-th]].
  %%CITATION = doi:10.1103/PhysRevLett.113.102301;%%
  %72 citations counted in INSPIRE as of 09 Apr 2017
  
        %\cite{Bjorken:2013boa}
\bibitem{Bjorken:2013boa} 
  J.~D.~Bjorken, S.~J.~Brodsky and A.~Scharff Goldhaber,
  %``Possible multiparticle ridge-like correlations in very high multiplicity proton-proton collisions,''
  Phys.\ Lett.\ B {\bf 726}, 344 (2013)
  doi:10.1016/j.physletb.2013.08.066
  [arXiv:1308.1435 [hep-ph]].
  %%CITATION = doi:10.1016/j.physletb.2013.08.066;%%
  %42 citations counted in INSPIRE as of 26 Jan 2018
  
  
  %\cite{Mantysaari:2017cni}
\bibitem{Mantysaari:2017cni} 
  H.~M\"{a}ntysaari, B.~Schenke, C.~Shen and P.~Tribedy,
  %``Imprints of fluctuating proton shapes on flow in proton-lead collisions at the LHC,''
  arXiv:1705.03177 [nucl-th].
  %%CITATION = ARXIV:1705.03177;%%
 
  
  %\cite{Mantysaari:2016ykx}
\bibitem{Mantysaari:2016ykx} 
  H.~M\"{a}ntysaari and B.~Schenke,
  %``Evidence of strong proton shape fluctuations from incoherent diffraction,''
  Phys.\ Rev.\ Lett.\  {\bf 117}, no. 5, 052301 (2016)
  doi:10.1103/PhysRevLett.117.052301
  [arXiv:1603.04349 [hep-ph]].
  %%CITATION = doi:10.1103/PhysRevLett.117.052301;%%
  %22 citations counted in INSPIRE as of 01 May 2017

%\cite{Greif:2017bnr}
\bibitem{Greif:2017bnr} 
  M.~Greif, C.~Greiner, B.~Schenke, S.~Schlichting and Z.~Xu,
  %``Importance of initial and final state effects for azimuthal correlations in p+Pb collisions,''
  Phys.\ Rev.\ D {\bf 96}, no. 9, 091504 (2017)
  doi:10.1103/PhysRevD.96.091504
  [arXiv:1708.02076 [hep-ph]].
  %%CITATION = doi:10.1103/PhysRevD.96.091504;%%
  %2 citations counted in INSPIRE as of 26 Jan 2018
  
  %\cite{Zhao:2017rgg}
\bibitem{Zhao:2017rgg} 
  W.~Zhao, Y.~Zhou, H.~j.~Xu, W.~Deng and H.~Song,
  %``Hydrodynamic Collectivity in Proton--Proton Collisions at 13 TeV,''
  arXiv:1801.00271 [nucl-th].
  %%CITATION = ARXIV:1801.00271;%%
  
  %\cite{Bozek:2013uha}
\bibitem{Bozek:2013uha} 
  P.~Bozek and W.~Broniowski,
  %``Collective dynamics in high-energy proton-nucleus collisions,''
  Phys.\ Rev.\ C {\bf 88}, no. 1, 014903 (2013)
  doi:10.1103/PhysRevC.88.014903
  [arXiv:1304.3044 [nucl-th]].
  %%CITATION = doi:10.1103/PhysRevC.88.014903;%%
  %165 citations counted in INSPIRE as of 26 Jan 2018
  
    %\cite{Kozlov:2014fqa}
\bibitem{Kozlov:2014fqa} 
  I.~Kozlov, M.~Luzum, G.~Denicol, S.~Jeon and C.~Gale,
  %``Transverse momentum structure of pair correlations as a signature of collective behavior in small collision systems,''
  arXiv:1405.3976 [nucl-th].
  %%CITATION = ARXIV:1405.3976;%%
  %53 citations counted in INSPIRE as of 26 Jun 2017
  
  %\cite{Kozlov:2014hya}
\bibitem{Kozlov:2014hya} 
  I.~Kozlov, M.~Luzum, G.~S.~Denicol, S.~Jeon and C.~Gale,
  %``Signatures of collective behavior in small systems,''
  Nucl.\ Phys.\ A {\bf 931}, 1045 (2014)
  doi:10.1016/j.nuclphysa.2014.09.054
  [arXiv:1412.3147 [nucl-th]].
  %%CITATION = doi:10.1016/j.nuclphysa.2014.09.054;%%
  %8 citations counted in INSPIRE as of 26 Jan 2018

  %\cite{Iancu:2003xm}
\bibitem{Iancu:2003xm} 
  E.~Iancu and R.~Venugopalan,
  %``The Color glass condensate and high-energy scattering in QCD,''
  In *Hwa, R.C. (ed.) et al.: Quark gluon plasma* 249-3363
  doi:10.1142/9789812795533\_0005
  [hep-ph/0303204].
  %%CITATION = doi:10.1142/9789812795533_0005;%%
  %677 citations counted in INSPIRE as of 09 Apr 2017

%\cite{Gelis:2010nm}
\bibitem{Gelis:2010nm} 
  F.~Gelis, E.~Iancu, J.~Jalilian-Marian and R.~Venugopalan,
  %``The Color Glass Condensate,''
  Ann.\ Rev.\ Nucl.\ Part.\ Sci.\  {\bf 60}, 463 (2010)
  doi:10.1146/annurev.nucl.010909.083629
  [arXiv:1002.0333 [hep-ph]].
  %%CITATION = doi:10.1146/annurev.nucl.010909.083629;%%
  %561 citations counted in INSPIRE as of 09 Apr 2017
  
      
  %\cite{Dumitru:2008wn}
\bibitem{Dumitru:2008wn} 
  A.~Dumitru, F.~Gelis, L.~McLerran and R.~Venugopalan,
  %``Glasma flux tubes and the near side ridge phenomenon at RHIC,''
  Nucl.\ Phys.\ A {\bf 810}, 91 (2008)
  doi:10.1016/j.nuclphysa.2008.06.012
  [arXiv:0804.3858 [hep-ph]].
  %%CITATION = doi:10.1016/j.nuclphysa.2008.06.012;%%
  %290 citations counted in INSPIRE as of 01 May 2017
  
  %\cite{Dusling:2009ar}
\bibitem{Dusling:2009ar} 
  K.~Dusling, D.~Fernandez-Fraile and R.~Venugopalan,
  %``Three-particle correlation from glasma flux tubes,''
  Nucl.\ Phys.\ A {\bf 828}, 161 (2009)
  doi:10.1016/j.nuclphysa.2009.06.017
  [arXiv:0902.4435 [nucl-th]].
  %%CITATION = doi:10.1016/j.nuclphysa.2009.06.017;%%
  %48 citations counted in INSPIRE as of 26 Jan 2018
  
  %\cite{Dumitru:2010iy}
\bibitem{Dumitru:2010iy} 
  A.~Dumitru, K.~Dusling, F.~Gelis, J.~Jalilian-Marian, T.~Lappi and R.~Venugopalan,
  %``The Ridge in proton-proton collisions at the LHC,''
  Phys.\ Lett.\ B {\bf 697}, 21 (2011)
  doi:10.1016/j.physletb.2011.01.024
  [arXiv:1009.5295 [hep-ph]].
  %%CITATION = doi:10.1016/j.physletb.2011.01.024;%%
  %189 citations counted in INSPIRE as of 09 Apr 2017
 
  
  %\cite{Dusling:2012iga}
\bibitem{Dusling:2012iga} 
  K.~Dusling and R.~Venugopalan,
  %``Azimuthal collimation of long range rapidity correlations by strong color fields in high multiplicity hadron-hadron collisions,''
  Phys.\ Rev.\ Lett.\  {\bf 108}, 262001 (2012)
  doi:10.1103/PhysRevLett.108.262001
  [arXiv:1201.2658 [hep-ph]].
  %%CITATION = doi:10.1103/PhysRevLett.108.262001;%%
  %124 citations counted in INSPIRE as of 09 Apr 2017
  
%\cite{Dusling:2012cg}
\bibitem{Dusling:2012cg} 
  K.~Dusling and R.~Venugopalan,
  %``Evidence for BFKL and saturation dynamics from dihadron spectra at the LHC,''
  Phys.\ Rev.\ D {\bf 87}, no. 5, 051502 (2013)
  doi:10.1103/PhysRevD.87.051502
  [arXiv:1210.3890 [hep-ph]].
  %%CITATION = doi:10.1103/PhysRevD.87.051502;%%
  %118 citations counted in INSPIRE as of 09 Apr 2017
  
  %\cite{Dusling:2012wy}
\bibitem{Dusling:2012wy} 
  K.~Dusling and R.~Venugopalan,
  %``Explanation of systematics of CMS p+Pb high multiplicity di-hadron data at $\sqrt{s}_{\rm NN} = 5.02$ TeV,''
  Phys.\ Rev.\ D {\bf 87}, no. 5, 054014 (2013)
  doi:10.1103/PhysRevD.87.054014
  [arXiv:1211.3701 [hep-ph]].
  %%CITATION = doi:10.1103/PhysRevD.87.054014;%%
  %155 citations counted in INSPIRE as of 09 Apr 2017
  
  %\cite{Dusling:2013qoz}
\bibitem{Dusling:2013qoz} 
  K.~Dusling and R.~Venugopalan,
  %``Comparison of the color glass condensate to dihadron correlations in proton-proton and proton-nucleus collisions,''
  Phys.\ Rev.\ D {\bf 87}, no. 9, 094034 (2013)
  doi:10.1103/PhysRevD.87.094034
  [arXiv:1302.7018 [hep-ph]].
  %%CITATION = doi:10.1103/PhysRevD.87.094034;%%
  %211 citations counted in INSPIRE as of 09 Apr 2017
  
    %\cite{Dusling:2015rja}
\bibitem{Dusling:2015rja} 
  K.~Dusling, P.~Tribedy and R.~Venugopalan,
  %``Energy dependence of the ridge in high multiplicity proton-proton collisions,''
  Phys.\ Rev.\ D {\bf 93}, no. 1, 014034 (2016)
  doi:10.1103/PhysRevD.93.014034
  [arXiv:1509.04410 [hep-ph]].
  %%CITATION = doi:10.1103/PhysRevD.93.014034;%%
  %17 citations counted in INSPIRE as of 09 Apr 2017
  
  %\cite{Ozonder:2017moj}
\bibitem{Ozonder:2017moj} 
  S. Oz\"{o}nder,
  %``Cumulant Expansion in Gluon Saturation, and Five and Six-Gluon Azimuthal Correlations,''
  Phys.\ Rev.\ D {\bf 96}, no. 7, 074005 (2017)
  doi:10.1103/PhysRevD.96.074005
  [arXiv:1710.00425 [hep-ph]].
  %%CITATION = doi:10.1103/PhysRevD.96.074005;%%
  %1 citations counted in INSPIRE as of 26 Jan 2018
  
          %\cite{McLerran:1993ka}
\bibitem{McLerran:1993ka} 
  L.~D.~McLerran and R.~Venugopalan,
  %``Gluon distribution functions for very large nuclei at small transverse momentum,''
  Phys.\ Rev.\ D {\bf 49}, 3352 (1994)
  doi:10.1103/PhysRevD.49.3352
  [hep-ph/9311205].
  %%CITATION = doi:10.1103/PhysRevD.49.3352;%%
  %1204 citations counted in INSPIRE as of 09 Apr 2017
  
%\cite{McLerran:1993ni}
\bibitem{McLerran:1993ni} 
  L.~D.~McLerran and R.~Venugopalan,
  %``Computing quark and gluon distribution functions for very large nuclei,''
  Phys.\ Rev.\ D {\bf 49}, 2233 (1994)
  doi:10.1103/PhysRevD.49.2233
  [hep-ph/9309289].
  %%CITATION = doi:10.1103/PhysRevD.49.2233;%%
  
  %\cite{Kovchegov:1996ty}
\bibitem{Kovchegov:1996ty} 
  Y.~V.~Kovchegov,
  %``NonAbelian Weizsacker-Williams field and a two-dimensional effective color charge density for a very large nucleus,''
  Phys.\ Rev.\ D {\bf 54}, 5463 (1996)
  doi:10.1103/PhysRevD.54.5463
  [hep-ph/9605446].
  %%CITATION = doi:10.1103/PhysRevD.54.5463;%%
  %418 citations counted in INSPIRE as of 26 Jan 2018
  
  %\cite{Krasnitz:1998ns}
\bibitem{Krasnitz:1998ns} 
  A.~Krasnitz and R.~Venugopalan,
  %``Nonperturbative computation of gluon minijet production in nuclear collisions at very high-energies,''
  Nucl.\ Phys.\ B {\bf 557}, 237 (1999)
  doi:10.1016/S0550-3213(99)00366-1
  [hep-ph/9809433].
  %%CITATION = doi:10.1016/S0550-3213(99)00366-1;%%
  %278 citations counted in INSPIRE as of 26 Jan 2018
  
  %\cite{Krasnitz:2001qu}
\bibitem{Krasnitz:2001qu} 
  A.~Krasnitz, Y.~Nara and R.~Venugopalan,
  %``Coherent gluon production in very high-energy heavy ion collisions,''
  Phys.\ Rev.\ Lett.\  {\bf 87}, 192302 (2001)
  doi:10.1103/PhysRevLett.87.192302
  [hep-ph/0108092].
  %%CITATION = doi:10.1103/PhysRevLett.87.192302;%%
  %238 citations counted in INSPIRE as of 26 Jan 2018
  
  %\cite{Krasnitz:2003jw}
\bibitem{Krasnitz:2003jw} 
  A.~Krasnitz, Y.~Nara and R.~Venugopalan,
  %``Classical gluodynamics of high-energy nuclear collisions: An Erratumn and an update,''
  Nucl.\ Phys.\ A {\bf 727}, 427 (2003)
  doi:10.1016/j.nuclphysa.2003.08.004
  [hep-ph/0305112].
  %%CITATION = doi:10.1016/j.nuclphysa.2003.08.004;%%
  %160 citations counted in INSPIRE as of 26 Jan 2018
  
  %\cite{Lappi:2003bi}
\bibitem{Lappi:2003bi} 
  T.~Lappi,
  %``Production of gluons in the classical field model for heavy ion collisions,''
  Phys.\ Rev.\ C {\bf 67}, 054903 (2003)
  doi:10.1103/PhysRevC.67.054903
  [hep-ph/0303076].
  %%CITATION = doi:10.1103/PhysRevC.67.054903;%%
  %267 citations counted in INSPIRE as of 26 Jan 2018
  
  %\cite{Lappi:2007ku}
\bibitem{Lappi:2007ku} 
  T.~Lappi,
  %``Wilson line correlator in the MV model: Relating the glasma to deep inelastic scattering,''
  Eur.\ Phys.\ J.\ C {\bf 55}, 285 (2008)
  doi:10.1140/epjc/s10052-008-0588-4
  [arXiv:0711.3039 [hep-ph]].
  %%CITATION = doi:10.1140/epjc/s10052-008-0588-4;%%
  %67 citations counted in INSPIRE as of 26 Jan 2018
  
          %\cite{Schenke:2015aqa}
\bibitem{Schenke:2015aqa} 
  B.~Schenke, S.~Schlichting and R.~Venugopalan,
  %``Azimuthal anisotropies in p$+$Pb collisions from classical Yang?Mills dynamics,''
  Phys.\ Lett.\ B {\bf 747}, 76 (2015)
  doi:10.1016/j.physletb.2015.05.051
  [arXiv:1502.01331 [hep-ph]].
  %%CITATION = doi:10.1016/j.physletb.2015.05.051;%%
  %45 citations counted in INSPIRE as of 09 Apr 2017
  
        %\cite{Schenke:2016lrs}
\bibitem{Schenke:2016lrs} 
  B.~Schenke, S.~Schlichting, P.~Tribedy and R.~Venugopalan,
  %``Mass ordering of spectra from fragmentation of saturated gluon states in high multiplicity proton-proton collisions,''
  Phys.\ Rev.\ Lett.\  {\bf 117}, no. 16, 162301 (2016)
  doi:10.1103/PhysRevLett.117.162301
  [arXiv:1607.02496 [hep-ph]].
  %%CITATION = doi:10.1103/PhysRevLett.117.162301;%%
  %14 citations counted in INSPIRE as of 09 Apr 2017
  
%\cite{Andersson:1983ia}
\bibitem{Andersson:1983ia} 
  B.~Andersson, G.~Gustafson, G.~Ingelman and T.~Sjostrand,
  %``Parton Fragmentation and String Dynamics,''
  Phys.\ Rept.\  {\bf 97}, 31 (1983).
  doi:10.1016/0370-1573(83)90080-7
  %%CITATION = doi:10.1016/0370-1573(83)90080-7;%%
  %3069 citations counted in INSPIRE as of 26 Jan 2018
  
  %\cite{Sjostrand:1982fn}
\bibitem{Sjostrand:1982fn} 
  T.~Sjostrand,
  %``The Lund Monte Carlo for Jet Fragmentation,''
  Comput.\ Phys.\ Commun.\  {\bf 27}, 243 (1982).
  doi:10.1016/0010-4655(82)90175-8
  %%CITATION = doi:10.1016/0010-4655(82)90175-8;%%
  %582 citations counted in INSPIRE as of 26 Jan 2018
  
      %\cite{McLerran:2016snu}
\bibitem{McLerran:2016snu} 
  L.~McLerran and V.~Skokov,
  %``Odd Azimuthal Anisotropy of the Glasma for pA Scattering,''
  Nucl.\ Phys.\ A {\bf 959}, 83 (2017)
  doi:10.1016/j.nuclphysa.2016.12.011
  [arXiv:1611.09870 [hep-ph]].
  %%CITATION = doi:10.1016/j.nuclphysa.2016.12.011;%%
  %6 citations counted in INSPIRE as of 30 Aug 2017
  
  \bibitem{Skokov} 
V.~Skokov, private communication.

  \bibitem{Mace-Skokov-Tribedy-RV}
  M.~Mace, V.~Skokov, P.~Tribedy, R.~Venugopalan, in progress.
  
    %\cite{Dusling:2017dqg}
\bibitem{Dusling:2017dqg} 
  K.~Dusling, M.~Mace and R.~Venugopalan,
  %``Multiparticle collectivity from initial state correlations in high energy proton-nucleus collisions,''
    Phys.\ Rev.\ Lett.\ {\bf 120}, 042002 (2018)
    doi:10.1103/PhysRevLett.120.042002
  [arXiv:1705.00745 [hep-ph]].
  %%CITATION = ARXIV:1705.00745;%%
  %11 citations counted in INSPIRE as of 27 Dec 2017
    
      %1649 citations counted in INSPIRE as of 09 Apr 2017
%\cite{Dusling:2017aot}
\bibitem{Dusling:2017aot} 
  K.~Dusling, M.~Mace and R.~Venugopalan,
  %``Parton model description of multiparticle azimuthal correlations in $pA$ collisions,''
    Phys.\ Rev.\ D {\bf 97}, 016014 (2018)
    doi:10.1103/PhysRevD.97.016014
  [arXiv:1706.06260 [hep-ph]].
  %%CITATION = ARXIV:1706.06260;%%  

    %\cite{Borghini:2001vi}
\bibitem{Borghini:2001vi} 
  N.~Borghini, P.~M.~Dinh and J.~Y.~Ollitrault,
  %``Flow analysis from multiparticle azimuthal correlations,''
  Phys.\ Rev.\ C {\bf 64}, 054901 (2001)
  doi:10.1103/PhysRevC.64.054901
  [nucl-th/0105040].
  %%CITATION = doi:10.1103/PhysRevC.64.054901;%%
  %313 citations counted in INSPIRE as of 09 Apr 2017
  
      
   %\cite{Bzdak:2013rya}
\bibitem{Bzdak:2013rya} 
  A.~Bzdak, P.~Bozek and L.~McLerran,
  %``Fluctuation induced equality of multi-particle eccentricities for four or more particles,''
  Nucl.\ Phys.\ A {\bf 927}, 15 (2014)
  doi:10.1016/j.nuclphysa.2014.03.007
  [arXiv:1311.7325 [hep-ph]].
  %%CITATION = doi:10.1016/j.nuclphysa.2014.03.007;%%
  %43 citations counted in INSPIRE as of 01 May 2017


  
      %\cite{Gardim:2011xv}
\bibitem{Gardim:2011xv} 
  F.~G.~Gardim, F.~Grassi, M.~Luzum and J.~Y.~Ollitrault,
  %``Mapping the hydrodynamic response to the initial geometry in heavy-ion collisions,''
  Phys.\ Rev.\ C {\bf 85}, 024908 (2012)
  doi:10.1103/PhysRevC.85.024908
  [arXiv:1111.6538 [nucl-th]].
  %%CITATION = doi:10.1103/PhysRevC.85.024908;%%
  %161 citations counted in INSPIRE as of 27 Dec 2017
  
  %\cite{Niemi:2012aj}
\bibitem{Niemi:2012aj} 
  H.~Niemi, G.~S.~Denicol, H.~Holopainen and P.~Huovinen,
  %``Event-by-event distributions of azimuthal asymmetries in ultrarelativistic heavy-ion collisions,''
  Phys.\ Rev.\ C {\bf 87}, no. 5, 054901 (2013)
  doi:10.1103/PhysRevC.87.054901
  [arXiv:1212.1008 [nucl-th]].
  %%CITATION = doi:10.1103/PhysRevC.87.054901;%%
  %143 citations counted in INSPIRE as of 27 Dec 2017
  
    
  %\cite{Blaizot:2014nia}
\bibitem{Blaizot:2014nia} 
  J.~P.~Blaizot, W.~Broniowski and J.~Y.~Ollitrault,
  %``Continuous description of fluctuating eccentricities,''
  Phys.\ Lett.\ B {\bf 738}, 166 (2014)
  doi:10.1016/j.physletb.2014.09.028
  [arXiv:1405.3572 [nucl-th]].
  %%CITATION = doi:10.1016/j.physletb.2014.09.028;%%
  %5 citations counted in INSPIRE as of 26 Jan 2018
  
%\cite{Gronqvist:2016hym}
\bibitem{Gronqvist:2016hym} 
  H.~Gr\"{o}nqvist, J.~P.~Blaizot and J.~Y.~Ollitrault,
  %``Non-Gaussian eccentricity fluctuations,''
  Phys.\ Rev.\ C {\bf 94}, no. 3, 034905 (2016)
  doi:10.1103/PhysRevC.94.034905
  [arXiv:1604.07230 [nucl-th]].
  %%CITATION = doi:10.1103/PhysRevC.94.034905;%%
  %11 citations counted in INSPIRE as of 26 Jan 2018
  
        %\cite{Lappi:2015vha}
\bibitem{Lappi:2015vha} 
  T.~Lappi,
  %``Azimuthal harmonics of color fields in a high energy nucleus,''
  Phys.\ Lett.\ B {\bf 744}, 315 (2015)
  doi:10.1016/j.physletb.2015.04.015
  [arXiv:1501.05505 [hep-ph]].
  %%CITATION = doi:10.1016/j.physletb.2015.04.015;%%
  %31 citations counted in INSPIRE as of 01 May 2017

  
    %\cite{Lappi:2015vta}
\bibitem{Lappi:2015vta} 
  T.~Lappi, B.~Schenke, S.~Schlichting and R.~Venugopalan,
  %``Tracing the origin of azimuthal gluon correlations in the color glass condensate,''
  JHEP {\bf 1601}, 061 (2016)
  doi:10.1007/JHEP01(2016)061
  [arXiv:1509.03499 [hep-ph]].
  %%CITATION = doi:10.1007/JHEP01(2016)061;%%
  %23 citations counted in INSPIRE as of 09 Apr 2017
  
     %\cite{Dumitru:2002qt}
\bibitem{Dumitru:2002qt} 
  A.~Dumitru and J.~Jalilian-Marian,
  %``Forward quark jets from protons shattering the colored glass,''
  Phys.\ Rev.\ Lett.\  {\bf 89}, 022301 (2002)
  doi:10.1103/PhysRevLett.89.022301
  [hep-ph/0204028].
  %%CITATION = doi:10.1103/PhysRevLett.89.022301;%%
  %153 citations counted in INSPIRE as of 02 Feb 2017

      %\cite{Kowalski:2006hc}
\bibitem{Kowalski:2006hc} 
  H.~Kowalski, L.~Motyka and G.~Watt,
  %``Exclusive diffractive processes at HERA within the dipole picture,''
  Phys.\ Rev.\ D {\bf 74}, 074016 (2006)
  doi:10.1103/PhysRevD.74.074016
  [hep-ph/0606272].
  %%CITATION = doi:10.1103/PhysRevD.74.074016;%%
  %291 citations counted in INSPIRE as of 11 Apr 2017
  
  %\cite{Rezaeian:2012ji}
\bibitem{Rezaeian:2012ji} 
  A.~H.~Rezaeian, M.~Siddikov, M.~Van de Klundert and R.~Venugopalan,
  %``Analysis of combined HERA data in the Impact-Parameter dependent Saturation model,''
  Phys.\ Rev.\ D {\bf 87}, no. 3, 034002 (2013)
  doi:10.1103/PhysRevD.87.034002
  [arXiv:1212.2974 [hep-ph]].
  %%CITATION = doi:10.1103/PhysRevD.87.034002;%%
  %105 citations counted in INSPIRE as of 26 Jan 2018
  
  %\cite{Gelis:2008ad}
\bibitem{Gelis:2008ad} 
  F.~Gelis, T.~Lappi and R.~Venugopalan,
  %``High energy factorization in nucleus-nucleus collisions. II. Multigluon correlations,''
  Phys.\ Rev.\ D {\bf 78}, 054020 (2008)
  doi:10.1103/PhysRevD.78.054020
  [arXiv:0807.1306 [hep-ph]].
  %%CITATION = doi:10.1103/PhysRevD.78.054020;%%
  %96 citations counted in INSPIRE as of 26 Jan 2018
  
      %\cite{Dusling:2009ni}
\bibitem{Dusling:2009ni} 
  K.~Dusling, F.~Gelis, T.~Lappi and R.~Venugopalan,
  %``Long range two-particle rapidity correlations in A+A collisions from high energy QCD evolution,''
  Nucl.\ Phys.\ A {\bf 836}, 159 (2010)
  doi:10.1016/j.nuclphysa.2009.12.044
  [arXiv:0911.2720 [hep-ph]].
  %%CITATION = doi:10.1016/j.nuclphysa.2009.12.044;%%
  %137 citations counted in INSPIRE as of 08 Jan 2018
  
  %\cite{Kovchegov:2012nd}
\bibitem{Kovchegov:2012nd} 
  Y.~V.~Kovchegov and D.~E.~Wertepny,
  %``Long-Range Rapidity Correlations in Heavy-Light Ion Collisions,''
  Nucl.\ Phys.\ A {\bf 906}, 50 (2013)
  doi:10.1016/j.nuclphysa.2013.03.006
  [arXiv:1212.1195 [hep-ph]].
  %%CITATION = doi:10.1016/j.nuclphysa.2013.03.006;%%
  %73 citations counted in INSPIRE as of 14 Dec 2017
  
  %\cite{Kovner:2017vro}
\bibitem{Kovner:2017vro} 
  A.~Kovner and A.~H.~Rezaeian,
  %``Double parton scattering in the color glass condensate: Hanbury-Brown-Twiss correlations in double inclusive photon production,''
  Phys.\ Rev.\ D {\bf 95}, no. 11, 114028 (2017)
  doi:10.1103/PhysRevD.95.114028
  [arXiv:1701.00494 [hep-ph]].
  %%CITATION = doi:10.1103/PhysRevD.95.114028;%%
  %5 citations counted in INSPIRE as of 26 Jan 2018

%\cite{Kovner:2018vec}
\bibitem{Kovner:2018vec} 
  A.~Kovner and A.~H.~Rezaeian,
  %``Multi Quark Production in p+A collisions: Quantum Interference Effects,''
  arXiv:1801.04875 [hep-ph].
  %%CITATION = ARXIV:1801.04875;%%
  
    
   \bibitem{Kovner:2001vi} 
  A.~Kovner and U.~A.~Wiedemann,
  %``Eikonal evolution and gluon radiation,''
  Phys.\ Rev.\ D {\bf 64}, 114002 (2001)
  doi:10.1103/PhysRevD.64.114002
  [hep-ph/0106240].
  %%CITATION = doi:10.1103/PhysRevD.64.114002;%%
  %124 citations counted in INSPIRE as of 10 May 2017
  
   %\cite{Blaizot:2004wu}
\bibitem{Blaizot:2004wu} 
  J.~P.~Blaizot, F.~Gelis and R.~Venugopalan,
  %``High-energy pA collisions in the color glass condensate approach. 1. Gluon production and the Cronin effect,''
  Nucl.\ Phys.\ A {\bf 743}, 13 (2004)
  doi:10.1016/j.nuclphysa.2004.07.005
  [hep-ph/0402256].
  %%CITATION = doi:10.1016/j.nuclphysa.2004.07.005;%%
  %175 citations counted in INSPIRE as of 11 Apr 2017
  
    %\cite{Dominguez:2008aa}
\bibitem{Dominguez:2008aa} 
  F.~Dominguez, C.~Marquet and B.~Wu,
  %``On multiple scatterings of mesons in hot and cold QCD matter,''
  Nucl.\ Phys.\ A {\bf 823}, 99 (2009)
  doi:10.1016/j.nuclphysa.2009.03.008
  [arXiv:0812.3878 [nucl-th]].
  %%CITATION = doi:10.1016/j.nuclphysa.2009.03.008;%%
  %20 citations counted in INSPIRE as of 02 Jul 2016

  
  %\cite{Fukushima:2007dy}
\bibitem{Fukushima:2007dy} 
  K.~Fukushima and Y.~Hidaka,
  %``Light projectile scattering off the color glass condensate,''
  JHEP {\bf 0706}, 040 (2007)
  doi:10.1088/1126-6708/2007/06/040
  [arXiv:0704.2806 [hep-ph]].
  %%CITATION = doi:10.1088/1126-6708/2007/06/040;%%
  %23 citations counted in INSPIRE as of 09 Jan 2018
  
  %\cite{Fukushima:2017mko}
\bibitem{Fukushima:2017mko} 
  K.~Fukushima and Y.~Hidaka,
  %``General formulae for dipole Wilson line correlators with the Color Glass Condensate,''
  JHEP {\bf 1711}, 114 (2017)
  doi:10.1007/JHEP11(2017)114
  [arXiv:1708.03051 [hep-ph]].
  %%CITATION = doi:10.1007/JHEP11(2017)114;%%
  
         
    %\cite{Dominguez:2012ad}
\bibitem{Dominguez:2012ad} 
  F.~Dominguez, C.~Marquet, A.~M.~Stasto and B.~W.~Xiao,
  %``Universality of multiparticle production in QCD at high energies,''
  Phys.\ Rev.\ D {\bf 87}, 034007 (2013)
  doi:10.1103/PhysRevD.87.034007
  [arXiv:1210.1141 [hep-ph]].
  %%CITATION = doi:10.1103/PhysRevD.87.034007;%%
  %28 citations counted in INSPIRE as of 24 Apr 2017
  
    
%\cite{Aaboud:2017blb}
\bibitem{Aaboud:2017blb} 
  M.~Aaboud {\it et al.} [ATLAS Collaboration],
  %``Measurement of multi-particle azimuthal correlations with the subevent cumulant method in $pp$ and $p$+Pb collisions with the ATLAS detector at the LHC,''
  arXiv:1708.03559 [hep-ex].
  %%CITATION = ARXIV:1708.03559;%%
  %8 citations counted in INSPIRE as of 08 Jan 2018
  
      %\cite{Gelis:2009wh}
\bibitem{Gelis:2009wh} 
  F.~Gelis, T.~Lappi and L.~McLerran,
  %``Glittering Glasmas,''
  Nucl.\ Phys.\ A {\bf 828}, 149 (2009)
  doi:10.1016/j.nuclphysa.2009.07.004
  [arXiv:0905.3234 [hep-ph]].
  %%CITATION = doi:10.1016/j.nuclphysa.2009.07.004;%%
  %99 citations counted in INSPIRE as of 08 Jan 2018
  
  %\cite{Basar:2013hea}
\bibitem{Basar:2013hea} 
  G.~Basar and D.~Teaney,
  %``Scaling relation between pA and AA collisions,''
  Phys.\ Rev.\ C {\bf 90}, no. 5, 054903 (2014)
  doi:10.1103/PhysRevC.90.054903
  [arXiv:1312.6770 [nucl-th]].
  %%CITATION = doi:10.1103/PhysRevC.90.054903;%%
  %45 citations counted in INSPIRE as of 26 Jan 2018
  
  %\cite{Adam:2015pya}
\bibitem{Adam:2015pya} 
  J.~Adam {\it et al.} [ALICE Collaboration],
  %``Two-pion femtoscopy in p-Pb collisions at $\sqrt{s_{\rm NN}}=5.02$ TeV,''
  Phys.\ Rev.\ C {\bf 91}, 034906 (2015)
  doi:10.1103/PhysRevC.91.034906
  [arXiv:1502.00559 [nucl-ex]].
  %%CITATION = doi:10.1103/PhysRevC.91.034906;%%
  %38 citations counted in INSPIRE as of 26 Jan 2018
  
    
    %\cite{Ball:2014uwa}
\bibitem{Ball:2014uwa} 
  R.~D.~Ball {\it et al.} [NNPDF Collaboration],
  %``Parton distributions for the LHC Run II,''
  JHEP {\bf 1504}, 040 (2015)
  doi:10.1007/JHEP04(2015)040
  [arXiv:1410.8849 [hep-ph]].
  %%CITATION = doi:10.1007/JHEP04(2015)040;%%
  %774 citations counted in INSPIRE as of 10 Aug 2017
  
  %\cite{Romatschke:2005pm}
\bibitem{Romatschke:2005pm} 
  P.~Romatschke and R.~Venugopalan,
  %``Collective non-Abelian instabilities in a melting color glass condensate,''
  Phys.\ Rev.\ Lett.\  {\bf 96}, 062302 (2006)
  doi:10.1103/PhysRevLett.96.062302
  [hep-ph/0510121].
  %%CITATION = doi:10.1103/PhysRevLett.96.062302;%%
  %225 citations counted in INSPIRE as of 26 Jan 2018
  
  %\cite{Romatschke:2006nk}
\bibitem{Romatschke:2006nk} 
  P.~Romatschke and R.~Venugopalan,
  %``The Unstable Glasma,''
  Phys.\ Rev.\ D {\bf 74}, 045011 (2006)
  doi:10.1103/PhysRevD.74.045011
  [hep-ph/0605045].
  %%CITATION = doi:10.1103/PhysRevD.74.045011;%%
  %209 citations counted in INSPIRE as of 26 Jan 2018
  
  %\cite{Berges:2013eia}
\bibitem{Berges:2013eia} 
  J.~Berges, K.~Boguslavski, S.~Schlichting and R.~Venugopalan,
  %``Turbulent thermalization process in heavy-ion collisions at ultrarelativistic energies,''
  Phys.\ Rev.\ D {\bf 89}, no. 7, 074011 (2014)
  doi:10.1103/PhysRevD.89.074011
  [arXiv:1303.5650 [hep-ph]].
  %%CITATION = doi:10.1103/PhysRevD.89.074011;%%
  %136 citations counted in INSPIRE as of 26 Jan 2018
  
  %\cite{Berges:2013fga}
\bibitem{Berges:2013fga} 
  J.~Berges, K.~Boguslavski, S.~Schlichting and R.~Venugopalan,
  %``Universal attractor in a highly occupied non-Abelian plasma,''
  Phys.\ Rev.\ D {\bf 89}, no. 11, 114007 (2014)
  doi:10.1103/PhysRevD.89.114007
  [arXiv:1311.3005 [hep-ph]].
  %%CITATION = doi:10.1103/PhysRevD.89.114007;%%
  %105 citations counted in INSPIRE as of 26 Jan 2018
  
  %\cite{Baier:2000sb}
\bibitem{Baier:2000sb} 
  R.~Baier, A.~H.~Mueller, D.~Schiff and D.~T.~Son,
  %``'Bottom up' thermalization in heavy ion collisions,''
  Phys.\ Lett.\ B {\bf 502}, 51 (2001)
  doi:10.1016/S0370-2693(01)00191-5
  [hep-ph/0009237].
  %%CITATION = doi:10.1016/S0370-2693(01)00191-5;%%
  %451 citations counted in INSPIRE as of 26 Jan 2018
  
      
\end{thebibliography}
\end{document}